\documentclass[prbl,twocolumn,showpacs,preprintnumbers]{revtex4-2}
\usepackage{amsmath}
\usepackage{amssymb}
\usepackage{amsfonts}
\usepackage{bbold}
\usepackage{mathrsfs}
\usepackage{graphicx, psfrag}
\usepackage[centerlast]{caption}
\usepackage{float}
\usepackage{subfig}
\usepackage{tikz}
\usepackage{physics}
\usepackage{anyfontsize}
\usepackage{pgfplots}
\usepackage[colorlinks=true, citecolor=blue, urlcolor = blue, linkcolor= red,bookmarks=true]{hyperref}
\usepackage{epstopdf}
\graphicspath{{new_figures/}}
\begin{document}
\newcommand{\ts}{\textsuperscript}
\def \beq{\begin{equation}}
\def \eeq{\end{equation}}
\def \bse{\begin{subequations}}
\def \ese{\end{subequations}}
\def \bea{\begin{eqnarray}}
\def \eea{\end{eqnarray}}
\def \bs{\boldsymbol}
\def \bem{\begin{displaymath}}
\def \eem{\end{displaymath}}
\def \bem{\begin{bmatrix}}
\def \eem{\end{bmatrix}}
\def \Ps{\hat{\Psi}(\boldsymbol{r})}
\def \bb{\bibitem}
\def \nn{\nonumber}
\def \bc{\begin{center}}
\def \ec{\end{center}}

\title{\textbf{Generation, manipulation and detection of snake state trajectories of a neutral atom in a ring-cavity}}

\author{Poornima Shakya$^{1}$, Nishant Dogra$^{2}$ and Sankalpa Ghosh$^{1}$}
\affiliation{${^1}$Department of Physics, Indian Institute of Technology Delhi, New Delhi, India\\
${^2}$Physics Department, Cavendish Laboratory, Cambridge, CB3 0HE, UK}

\begin{abstract}
We propose a set-up to create and detect the atomic counterpart of snake state trajectories which occur at the interface where the magnetic field reverses direction. Such a magnetic field is generated by coupling two counter-propagating modes of a ring cavity to a two-level atom. The spatial distribution and the strength of the induced magnetic field are controlled by the transverse mode profile of the cavity modes and the number of photons in the two modes, respectively. By analysing the atomic motion in such a magnetic field while including the cavity back-action, we find that the atom follows snake state trajectories which can be non-destructively detected and reconstructed from the phase and the intensity of the light field leaking from the cavity. We finally show that the system parameters can be tuned to modify the transport properties of the snake states and even amplify the effect of cavity feedback which can completely alter their topology.
\end{abstract}

\maketitle
\section{Introduction}
Laser-induced synthetic gauge fields, realized by coupling different internal atomic states \cite{Dalibard, Goldman, Spielman, Gerbier, Jaksch, Sorensen, Kolovsky, Cooper, JStruck, MAidels,  PHauke, JStruckPHauke, Miyake, MAtala, Lin_spielman, LinCompton, Lin_Garcia, Feder}, have provided a unique tool to the extremely well-controlled and tunable ultracold atomic systems \cite{Bloch, Blochrev, Buluta, Schmeidmayer}. Such gauge fields have paved the way for the realization and investigation of phenomena like quantum Hall effect \cite{Cooper2008, Viefers, Fetter}, spin-orbit coupling \cite{Lin_spielman, LinCompton, Lin_Garcia, Feder} and topological superfluidity \cite{Goldman, Ruhman, Jiang} as well as quantum simulation of fundamental topological models like Hofstadter model \cite{Jaksch, AidelsGold, Miyake, MAtala} and Haldane model \cite{Jotzu}. As opposed to static gauge fields, which are described as externally imposed potentials in the atomic Hamiltonian, the dynamical gauge fields additionally include the feedback from atomic dynamics and are a crucial ingredient of many fundamental gauge theories \cite{Kapit_Mueller, Zoller_Ban, Wiese, Zoller_Dal, Lewenstein_Tag, Zohar_Reznik}. One way to generate such feedback is to couple an ultracold atomic system to a single- or multi-mode optical cavity where the atomic wavefunction and its dynamics affect the phase and the intensity of the intracavity field, and in turn, the cavity field provides dynamical feedback on the atomic state. 
Such systems have been used to study phenomena such as Dicke superradiance in single pump \cite{Baumann} and two-pump \cite{PShakya} system, continuous supersolidity \cite{Leonard1}, dynamical spin-orbit coupling \cite{Kroeze}, and self-oscillating topological pump \cite{Dreon}, and are predicted to generate artificial Meissner effect \cite{Ballantine}, quantum magnetism \cite{Padhi2014, Mivehvar2019}, topological superradiant states \cite{Pan, Zheng,Keeling2014, Piazza2014, Chen2014} and self-organized chiral edge states \cite{Brenn_Kollath, Sheikh_Kollath, Wolff_Sheikh, Sheikh_Brenn, Halati_Sheikh, Halati_Kollath} (see \cite{RitschRev, RitschRev2} for a complete list).

An important motivation to realize synthetic gauge fields is to create topologically nontrivial quantum phases which support edge modes \cite{Pan,Brenn_Kollath} that are resilient to scattering from defects and disorders and can be useful for topological quantum computation \cite{Nayak2008,Stern2013}. One-dimensional snake trajectories that occur at the boundaries separating different magnetic or charge domains provide a convenient way of realising such protected modes in solid state electronic systems \cite{Mueller, Peeters, Peeters1, Nogaret, Cserti, Egger, Sim, Nogaret1, Puja, Pdye, Marcus, Rickhaus}. It has been noted in \cite{Cserti, Egger, Sim, Nogaret1, Puja} that such current-carrying magnetic edge states can couple differently with the current-carrying electrostatic edge states in the quantum Hall regime and change the conductivity. Such states have been experimentally observed in two-dimensional electron gas \cite{Pdye} and graphene \cite{Marcus, Rickhaus} by measuring transport properties such as current and conductance. However, the real-time detection of such states is difficult in condensed matter experiments and is crucial for their complete characterization to understand their role in conductivity enhancement in electronic systems \cite{Watanabe, Nogaret2, Hara, Solimany, Sim1, Matulis, Kim}.

In this paper, we theoretically demonstrate that atomic analogue of such snake states can be realized using an atom-cavity coupled system in a more efficient and versatile way than their electronic counterpart. We consider a two-level atom coupled to two counter-propagating and orthogonally-polarized running wave modes of a high-finesse ring cavity. Using a dressed-state approach, we show that a non-uniform synthetic magnetic field, with strength proportional to the difference in the photon number in the two cavity modes, can be generated. The spatial structure of this magnetic field is governed by the transverse mode profile of the cavity modes, and it changes its sign about a point of symmetry for a Gaussian mode profile. By solving semi-classical equations of motion of the system in the presence of such a magnetic field, we show that the atom follows a snake state trajectory. The presence of the cavity not only adds a dynamic character to the generated artificial magnetic field, but an analysis of the cavity transmission spectrum also allows real time monitoring of the atomic state in a non-destructive way. Here, we show that the phase and intensity of the light in the two cavity modes can be used to reconstruct the snake state trajectory in real time, and with minimal effect of the cavity back-action \cite{Stamper-Kurn1}. This is one of the key results in this manuscript. We further illustrate that we can manipulate the conductance properties of the snake states (amplitude and direction) by tuning the initial atomic speed orthogonal to the direction of transport, external pumping strength of the two cavity modes and the strength of atom-cavity coupling. Finally, we show that the effect of cavity back-action can be enhanced by making the (average) photon number in the two cavity modes comparable, which leads to the destruction of the topology of the snake states and results in the formation of states with more complex spatial trajectories. Our work will pave the way for long-distance transport in atomtronics via snake states  which can have technological applications in the fields of  quantum computation and quantum information processing \cite{ Folman, Holland, Amico, DJaksch}.

\begin{figure}
\centering
\includegraphics[width=1\columnwidth, height=1 \columnwidth] {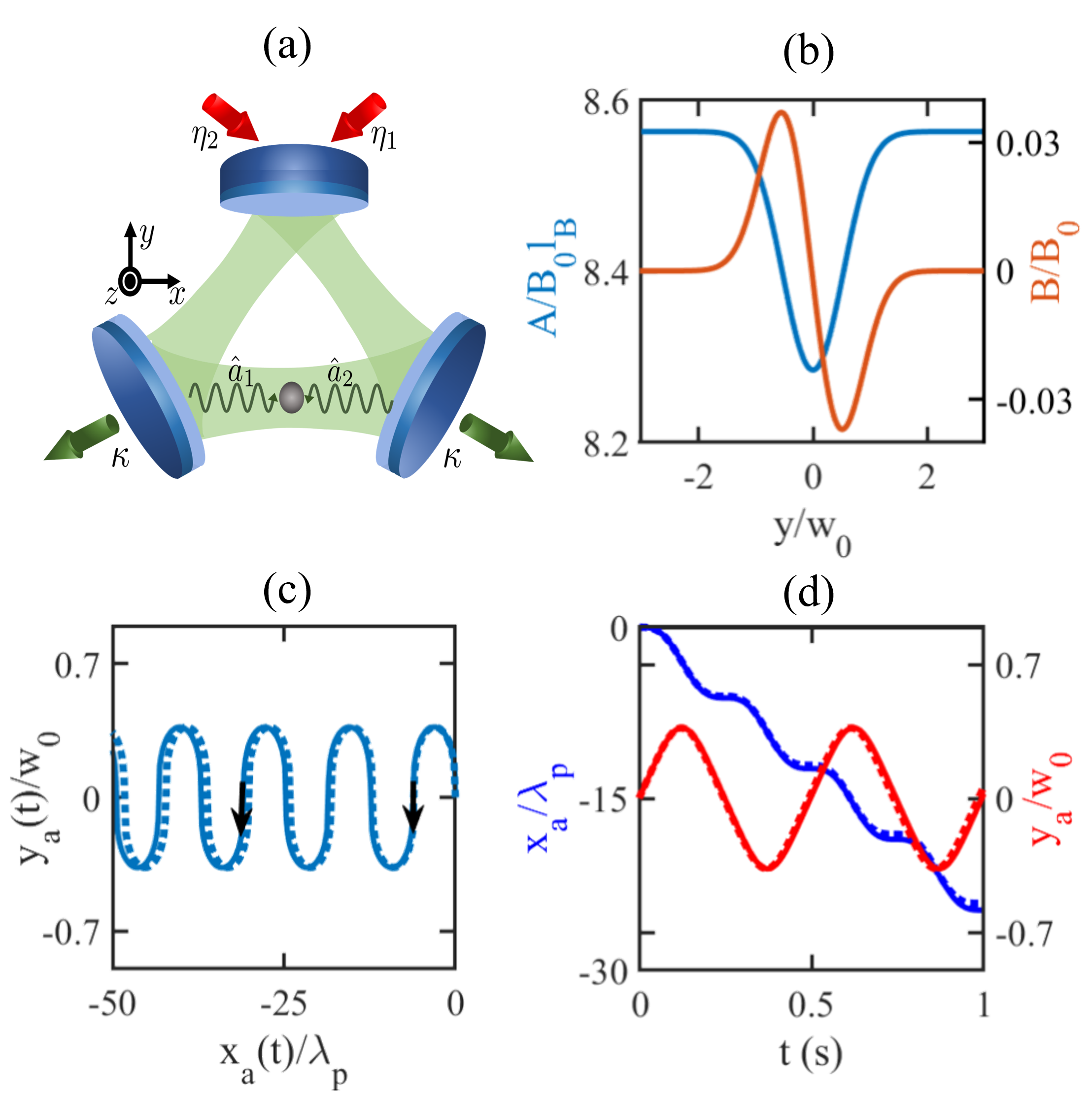}
\caption{({\it color online}) (a) Schematic of a single two-level atom trapped inside a ring cavity with two counter-propagating running wave modes which are orthogonally polarized and are described by the photon annihilation operators $\hat{a}_1$ and $\hat{a}_2$. The two modes are respectively pumped with strengths $\eta_1$ and $\eta_2$. $\kappa$ is the cavity decay rate. For (b-d) $\eta_1 \approx 2\pi\times 52$ MHz and $\eta_2=0$ which gives time-averaged photon numbers in the two cavity modes to be $\langle n_1\rangle\approx$ 342 and $\langle n_2\rangle\approx$ 8. (b) Vector potential $A$ ($blue$ curve), and magnetic field $B$ ($orange$ curve), as a function of $y$. $B_0$ and $l_B$ are respective natural magnetic field and magnetic length scales of the system, and $w_0$ is the waist size of the cavity modes, see text. (c) The snake state trajectory of the atom in the $x$-$y$ plane with $\lambda_p$ being the pump wavelength. The $black$ arrows indicate the direction of increasing time. (d) The $x$-position, $x_a$ ($blue$ curve)  and $y$-position,  $y_a$ ($red$ curve) of the atom as a function of time $t$. The dotted curves in (c) and (d) show the atomic trajectory, $x$-position and the $y$-position for fixed photon number $\langle n_1\rangle$ and $\langle n_2\rangle$ in the two cavity modes and thus excluding cavity back-action.}\label{Fig1}
\end{figure}

\section{System Hamiltonian}\label{SHEq}
We consider a single two-level atom with internal states $\vert g\rangle$ and $\vert e\rangle$ coupled to two counter-propagating running wave modes of a ring cavity as shown in Fig.~\ref{Fig1}(a). The two cavity modes are orthogonally polarized and are pumped on-axis with pump strengths $\eta_1$ and $\eta_2$. The total Hamiltonian describing the coupled atom-cavity system in the rotating frame of the pump field space \cite{Jaynes} can be written as (see Appendix \ref{SH} for details)-
\begin{eqnarray}
\hat{H}_{RF} &=& \hat{H}_{0}+\hat{H}_{I}\label{H_RF}
\end{eqnarray} 
where 
\bea 
\hat{H}_0 & = & \frac{\hat{P}^2}{2m_a}\hat{\mathbb{I}}\nn \\
\hat{H}_{I} &=& -\frac{\hbar \Delta_a\hat{\sigma}_z}{2} -\hbar\Delta_c\left(\hat{a}^{\dag}_{1}\hat{a}_{1}+\hat{a}^{\dag}_{2}\hat{a}_2\right)\nn\\
&+&\hbar\eta_1\left(\hat{a}_1 + \hat{a}^{\dag}_{1}\right) + \hbar\eta_2\left(\hat{a}_2 + \hat{a}^{\dag}_{2}\right)\nn\\
&+& \hbar  \left(g_1(y)\hat{\sigma}^{+}\hat{a}_1 e^{ikx} +g_2(y)\hat{\sigma}^{+}\hat{a}_2e^{-ikx}  \right.\nn \\
& +  & \left. g_1(y)\hat{\sigma}^{-}\hat{a}^{\dag}_1e^{-ikx}+g_2(y)\hat{\sigma}^{-}\hat{a}^{\dag}_2e^{ikx} \right).\nn
\eea 
$\hat{H}_0$ represents the kinetic energy of the atom, where $\hat{\mathbb{I}} = \vert g \rangle \langle g \vert + \vert e \rangle \langle e \vert$ is the identity operator in the internal two-dimensional Hilbert space of the atom, $m_a$ is the atom mass, and $\vec{r}$ and $\vec{P}$ are the atomic center-of-mass co-ordinate and momentum respectively. $\hat{H}_{I}$ represents the interaction Hamiltonian of the system, where $\omega_a$ is the atomic resonance frequency, $\omega_p$ is the pump frequency, $\omega_c$ is the cavity resonance frequency, $\Delta_a =\omega_p-\omega_a$ is the atom-pump detuning and $\Delta_{c}=\omega_p-\omega_{c}$ is the cavity-pump detuning. $\hat{\sigma}_z = \vert e\rangle\langle e\vert - \vert g\rangle\langle g\vert$ is the Pauli matrix and $\hat{\sigma}^{+}$ and $\hat{\sigma}^{-}$ are the atomic raising and lowering operators. $g_{1(2)}(y)= g_{10(20)}e^{-y^2/w^2_0} $ is the atom-photon coupling, with $w_0$ being the waist of the the two cavity modes and $g_{10(20)}=\frac{-\vec{d}\cdot\hat{e}_{y(z)}}{\hbar}\sqrt{\frac{\hbar \omega_{c}}{2\epsilon_0 V}}$ with $\vert g_{10}\vert=\vert g_{20}\vert = g_0$. $\hat{a}_1$ and $\hat{a}_2$ are the annihilation operators for the two cavity modes with respective spatial mode profiles of the form $e^{ikx}e^{-y^2/w_0^2}$ and $e^{-ikx}e^{-y^2/w_0^2}$, with $k=2\pi/\lambda_p$. Following a mean-field approach, we assume that the cavity fields can be described by a coherent state of the form $\vert \alpha_1,\alpha_2\rangle = \vert \alpha_1\rangle\vert \alpha_2\rangle$ with $\hat{a}_{1(2)}\vert \alpha_{1(2)}\rangle = \alpha_{1(2)}\vert \alpha_{1(2)}\rangle$ and $n_{1(2)} = \vert\alpha_{1(2)}\vert^2$ being the average photon number in the cavity mode 1(2). The phase associated with the cavity field in mode 1(2) is $\alpha_{1(2)}=\sqrt{n_{1(2)}}e^{i\phi_{1(2)}}$. We assume that all the dynamics take place in the $x$-$y$ plane, so we have neglected the $z$-coordinate in different expressions.

We diagonalise the interaction Hamiltonian $\hat{H}_{I}$ in the space spanned by the atom-photon bare-states, namely $\vert e,\alpha_1,\alpha_2\rangle, $ and $\vert g,\alpha_1,\alpha_2\rangle$, and obtain the eigenstates that are called dressed states $\vert D_{1(2)}\rangle$, for the coupled atom-photon system, with eigen energies $E_{1(2)}$, see Appendix \ref{SH} for details. Under adiabatic approximation \cite{Dalibard, Born,Truhlar1,Berry,Wilczek,Moody,Mead,Shapere,Sun,Sun1,Haroche}, we limit the system dynamics in the eigenspace of the lowest energy dressed state $\vert D_1\rangle$, and obtain the following equation for the evolution of the corresponding wave function $\psi_1$ (see Appendix \ref{SH}) -

\begin{widetext}
\beq
i\hbar \frac{\partial}{\partial t}\psi_{1}(\vec{r},t) = H_S\psi_1(\vec{r},t)=\Big[\frac{1}{2m_a}\Big\lbrace \left(\vec{p} - \vec{\textbf{A}}_{1,1}\right)^2 + \vert\vec{\textbf{A}}_{2,1}\vert^2  \Big\rbrace +E_1\Big]\psi_1(\vec{r},t)\label{EOM}
\eeq
\end{widetext}
Here, $\vec{\textbf{A}}_{1,1}$ acts as a synthetic vector potential while $\vec{\textbf{A}}_{2,1}$ contributes to the synthetic scalar potential term, given by  $W=\frac{1}{2m_a}\vert\vec{\textbf{A}}_{2,1}\vert^2 $. The last term of Eq.~(\ref{EOM}), $E_1$, acts as a deep trapping potential for the atomic centre-of-mass motion. In the next section, we will explain why the important dynamics of the system are governed only by the vector potential, $\vec{\textbf{A}}_{1,1}$. The scalar potential, the vector potential and the corresponding magnetic field obtained here depend on the difference in the photon number in the two cavity modes which dynamically depends on the position of the atom. 
 
The full expression for the cavity-induced synthetic vector potential in Eq.~(\ref{EOM}) is given as -
\begin{eqnarray}
\vec{\textbf{A}}_{1,1} &=& i\hbar \langle D_1 \vert \vec{\nabla}\vert D_1\rangle =A_x(y)\hat{x}\nn\\
&=& \frac{2\hbar k g^2(y)(n_1- n_2)}{G(G+\Delta_a)}\hat{x}
\end{eqnarray}
where 
$G = \sqrt{\Delta_a^2 + 4g^2(y)\left(n_1 +n_2 \right)}$. The corresponding synthetic magnetic field is -
\begin{eqnarray}
\vec{B} &=& -\frac{\partial A_x}{\partial y}\hat{z}= B_0 \frac{y}{w_0}\frac{\Delta_a}{G^3} 4g^2(y)( n_1 -n_2)\hat{z}\label{BFormula}
\end{eqnarray}
where $B_0=\frac{\hbar k}{w_0}$ defines the natural scale of the synthetic magnetic field with dimensions $[MT^{-1}]$ and the corresponding synthetic magnetic length is given by
$l_B=\sqrt{\frac{\hbar}{B_0}}$. Here, we observe that both the vector potential and the magnetic field scale with the difference in the photon numbers in the two cavity modes, $n_1$ and $n_2$ and thus can be tuned via $\eta_{1(2)}$, $\Delta_c$ and $g_{0}$. 
The expression for the scalar potential, $W$, is
\begin{eqnarray}
W = \frac{\hbar^2 k^2}{2m_a}(G^2-\Delta_a^2) \left[\left(\frac{y\Delta_a}{kw_0^2G^2}\right)^2 +\frac{1}{4G^2}\right]
\end{eqnarray}
where $E_R=\frac{\hbar^2 k^2}{2m_a}$ is the recoil energy of the atom.
We provide the spatial variation of the scalar potential in Appendix \ref{appen_scalar}.
The parameters considered in this work are for ${^{87}}{Rb}$ - $m_a =1.4\times 10^{-25}$ kg, $\kappa= 2\pi\times 650$ kHz, $\Delta_c= -5\kappa$, $g_{0}=2\pi\times 50$ MHz, $\lambda_p =780.25 $ nm, $\Delta_a \approx -2\pi\times 4.9$ GHz , $w_0 = 10$ $\mu$m, $\eta_{1} = 80\kappa$ and $\eta_2 = 0$. $\Gamma=2\pi\times 6$ MHz is the spontaneous emission rate of the atom. For these parameters, we get, $B_0 = 8.48\times 10^{-23}$ kg/s and $l_B=1.1$ $\mu$m. Using the magnetic length, the natural velocity scale of the system is $v_0=\hbar/(m_al_B) = 655$ $\mu$m/s. As $v_0 << \hbar k/m_a=5.9$ mm/s, the adiabatic approximation, which implies that when the
atom moves slowly enough, it remains in the state in which it started ($\vert D_1\rangle$ in our case), is justified. The presence of any external trap does not impact the shape of the resulting magnetic field; see Appendix \ref{appen_exttrap} for details.

In Fig.~\ref{Fig1}(b), we plot the vector potential and the magnetic field. The vector potential $A_x(y)$ has a symmetric Gaussian profile given by the cavity mode shape. The corresponding magnetic field $B_z$ scales linearly for $\vert y\vert<<w_0$ with a slope proportional to $n_1-n_2$ and reverses its direction about $y=0$. $B_z$ achieves its maximum magnitude at $\vert y\vert =0.5w_0$ and decays smoothly to $0$ for $\vert y\vert >0.5w_0$.
In the subsequent sections, we discuss the dynamics
of a single atom in the presence of such non-uniform magnetic fields
using a semi-classical
method. It may be pointed out that such inhomogeneous synthetic gauge field can
be created using different methods \cite{Lembessis,Sacha}. However, coupling to a ring-cavity allows for nearly non-destructive monitoring of the resultant dynamics as we show below.

\begin{figure*}
\centering
\includegraphics[width=2\columnwidth, height=0.55\columnwidth]{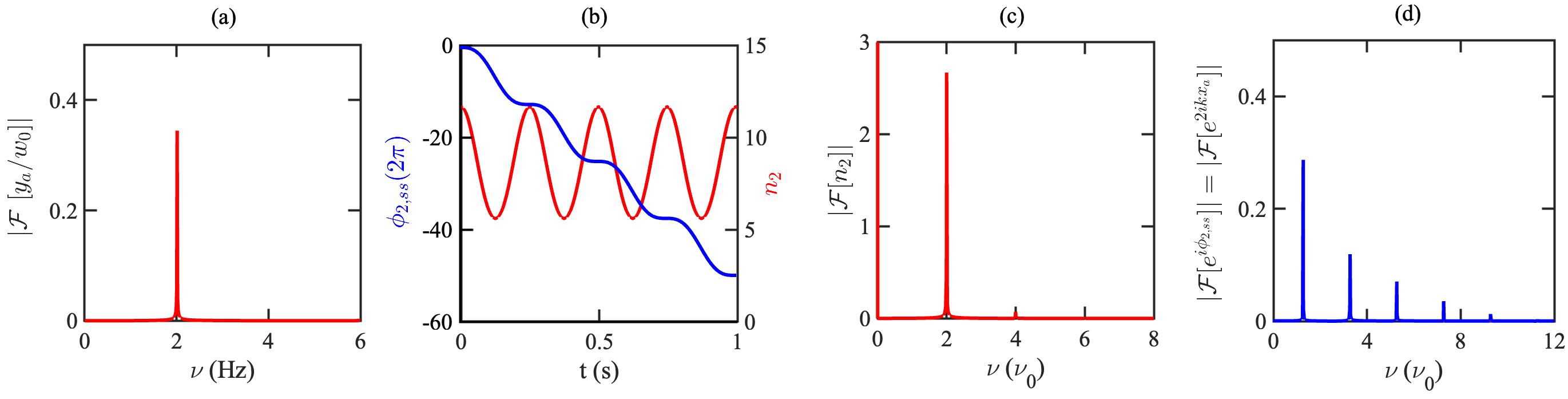}
\caption{\textit{(color online)}: (a)  Fourier transform for $y_a/w_0$ which shows a peak at $\nu_0 \approx 2$ Hz, (b) Photon number $n_2$ ($red$ curve) and the corresponding phase  $\phi_{2,ss}$ ($blue$ curve) in cavity mode 2 as a function of time $t$. (c) Fourier transform of $n_2$ which shows peaks at $2\nu_0$, $4\nu_0$, $6\nu_0$ and so on. (d) Fourier transform of $e^{i\phi_{2,ss}}$ which overlaps with the Fourier transform of $e^{2ikx_a}$.} \label{Fig2}
\end{figure*}

\section{Trajectories of a single atom and back-action of the cavity fields}\label{trajectories}
We now obtain the following semi-classical equation of motion for the atom due to the adiabatic following of the lowest energy dressed state \cite{Cheneau, Landau} -
\begin{equation}
m_a \frac{d \vec{v}}{dt}=-\vec{\nabla}E_{1} -\vec{\nabla}W(\vec{r}) +\vec{v}\times \vec{B}(\vec{r}). \label{eom_final}
\end{equation}
We want to isolate the effect of the $B(y)$ term on the atomic trajectory, so we neglect the $E_1+W$ contribution. In a multi-level atom, the effect of $E_1+W$ can be eliminated by proper choice of the pump wavelength. 
So, the two components of the equation of motion become
\begin{subequations}
\begin{eqnarray}
m_a \frac{d^2 x}{dt^2} &=& B(y) \frac{d y}{dt}\label{eom_vx_f}\\
m_a \frac{d^2 y}{dt^2} &=& -B(y) \frac{d x}{dt} \label{eom_vy_f}
\end{eqnarray}
\end{subequations}
The solution of the above equations gives us the atomic trajectory. To find the magnetic field, we need to additionally evaluate the number of photons in the two cavity modes which depend on the atomic position itself. As $\kappa >> \hbar k^2/2m_a, \hbar/(2m_al_b^2)$, we assume that the cavity field is always in a steady state and adiabatically follows the atomic motion. We, thus obtain the following expressions for the two cavity fields. 
\begin{eqnarray}
\alpha_{1(2)} &=&\langle \hat{a}_{1(2)} \rangle = \langle D_1 \vert \hat{a}_{1(2)} \vert D_1\rangle\nn\\
&=& \frac{i\eta_{1(2)}(i\bar{\Delta}_{c}-\bar{\kappa}) + i\eta_{2(1)}(iU+\gamma)e^{\mp 2ikx_a(t)}}{(i\bar{\Delta}_{c}-\bar{\kappa})^2-(iU+\gamma)^2} \label{alpha12}
\end{eqnarray}
where $U_{0}=\frac{\Delta_{a}g_{0}^2}{\Delta_{a}^2+4\Gamma^2}\approx \frac{g_0^2}{\Delta_a} \approx 2\pi\times 500$ kHz, $U=U_{0}e^{-2y_a^2(t)/w_0^2}$, $\gamma_0=\frac{2\Gamma g_{0}^2}{\Delta_{a}^2+4\Gamma^2}\approx \frac{2\Gamma g_0^2}{\Delta_a^2} \approx 2\pi\times 1.2$ kHz, $\gamma=\gamma_{0}e^{-2y_a^2(t)/w_0^2}$, $\bar{\Delta}_{c}=\Delta_{c}-U$, $\bar{\kappa}=\kappa+\gamma$ (the numerical values correspond to the parameters noted in the previous section).
To obtain the atomic trajectory in the $x$-$y$ plane, we solve Eq.~(\ref{eom_vx_f}, \ref{eom_vy_f}) simultaneously with Eq.~(\ref{alpha12}). We take initial velocities $v_{x0} = 0$ and $v_{y0} = 0.06v_0$ and initial positions: $x_{a0}=0$ and $y_{a0}=0$. We plot $y_a(t)$ as a function of $x_a(t)$ in Fig.~\ref{Fig1}(c) and realize that the particle drifts in the $-x$-direction while oscillating in the $y$-direction which is a snake state trajectory. The origin of such a trajectory is the following: The atom experiences a magnetic field having a finite slope which reverses its direction around $y=0$, and thus for $v_{x0} = 0 \neq v_{y0}$, a finite particle current is generated along $-x$-direction \cite{Mueller, Jackson}. The instantaneous radius of curvature $r(y)$ for the particle trajectory in a synthetic magnetic field is inversely proportional to the strength of the magnetic field: $r(y)\propto \frac{1}{B(y)}$. Therefore, for a large(small) magnitude of the magnetic field and thus large(small) $|y|$, the particle will trace a trajectory with a small(large) radius of curvature resulting in the peculiar snake state trajectory in the $x$-$y$ plane \cite{Mueller, Jackson}. The quantum fluctuations of the motion could give rise to a velocity distribution resulting in broader snake trajectories. However, with the presence of cavity noise and the finite lifetime of the excited state, such effects might be suppressed.  

We plot $x_a(t)$ and $y_a(t)$ as a function of $t$ in Fig.~\ref{Fig1}(d). Along $y$-direction, the particle oscillates periodically about $y=0$ at a fixed frequency as illustrated by the corresponding Fourier transform (illustrated by the symbol $\mathcal{F}$) in Fig.~\ref{Fig2}(a). Along $x$-direction, the particle has a finite average speed $\langle v_x \rangle$ where the notation $\langle S\rangle$ implies time-averaged value of a periodically oscillating quantity $S$. The dashed curves in Fig.~\ref{Fig1}(c, d) illustrate the particle trajectories when the number of photons in the two cavity modes is fixed to the time-averaged values as given by Eq.~(\ref{alpha12}); see Fig.~\ref{Fig2} for the full time evolution of the cavity fields. We observe that the cavity-feedback minimally affects the snake state trajectory of the particle and in general, this is true when $\langle n_1\rangle$ and $\langle n_2\rangle$ are very different from each other as we illustrate in Section \ref{breakdown}. The analytical formula for the period of the $y-$ trajectory can be estimated as
\begin{eqnarray}
y_{period}=\frac{1}{4\pi}\sqrt{\frac{m_a}{B_0v_{y0}}}\label{yperiod}
\end{eqnarray}
The above formula is obtained by approximating the Gaussian magnetic field by a linear spatially varying field for $y_a<w_0$ (see Fig.~\ref{Fig1}(b)). This is valid for $y_a^{pp}\leq w_0$ and for $y_a^{pp}\geq w_0$, the linear approximation of the Gaussian field breaks down. This gives us an idea about the parameters that affect the period of the snake trajectory and hence the frequency of the emitted photons. The pump wavelength, $\lambda_p$, the cavity waist mode, $w_0$ and the initial velocity of the atom along the $y-$ direction, $v_{y0}$, can be used to control the period of the trajectory. For small initial velocities, the change in the output photon number is very small and hence the detection of the trajectories will be difficult. The magnetic field dependence in the period also brings into the picture the laser fluctuations and the fluctuations in the output photon number $n_1$ and $n_2$ as can be seen in Eq.~(\ref{BFormula}). We will discuss the effect of the velocity and laser fluctuations in Section \ref{manipulation} where we calculate the signal-to-noise ratio as a function of initial velocity and pump strength. However, since the equations of motion do not have an exact analytical solution, the numerically obtained period of the oscillating trajectory can vary from the value obtained from the above formula. In Appendix \ref{freq_vel}, we show the numerical plot for the frequency of the snake trajectory in Fig.~\ref{Fig3Appen}. We also plot the frequency obtained from the $y_{period}$ in Eq.~(\ref{yperiod}) in Fig.~\ref{Fig3Appen}(a). 

\section{Cavity-based detection of the snake-like trajectories}\label{Cavity_detection}
To probe the snake state trajectory, we now look at the time evolution of the cavity field of mode 2, $\alpha_2 = \sqrt{n_2}e^{i\phi_{2,ss}}$, where the subscript $ss$ in $\phi_{2,ss}$ denotes the snake state phase. This is illustrated in Fig.~\ref{Fig2}(b). We realize that the time variation of phase $\phi_{2,ss}$ and photon number $n_2$ is qualitatively similar to $x_a$ and $y_a$, respectively. We can understand such behaviour as follows: The atom moving in the -$x$ direction absorbs a photon from cavity mode 1 and emits it into the counter-propagating mode (cavity mode 2). This decreases the atom's momentum by $2\hbar k$, and correspondingly, $e^{i2kx_a}$ phase is imprinted on the photon scattered into cavity mode 2, thus mapping $x_a$ on phase $\phi_{2,ss}$. The periodic atomic oscillation along the $y$-direction modulates the atom-cavity coupling $g(y)$, which has a Gaussian form (centered at $y=0$), and thus there is a periodic oscillation of $n_2$, which links $y_a$ to $n_2$. This can be easily seen via Eq.~(\ref{alpha12}) by assuming $|\Delta_{c}|\gg (\kappa, U_0) \gg \gamma_0$ (which is true for our case) where we obtain
\begin{eqnarray}
\alpha_2 \approx \frac{-\eta_1g^2_0e^{-2y_a^2(t)/w_0^2}e^{2ikx_a(t)}}{\Delta_a(i\bar{\Delta}_c-\bar{\kappa})^2},
\label{alpha2} 
\end{eqnarray}
which shows that $n_2 \propto e^{-4y_a^2(t)/w_0^2}$ and $\phi_{2,ss} \approx 2kx_a(t)$ (excluding constants coming from other prefactors).
We would like to point out that the above expression is obtained while ignoring the Doppler shift and more details on this aspect are provided in the later part of this section.
For a quantitative comparison, we look at the Fourier transform of various quantities. The Fourier transform of $y_a$ in Fig.~\ref{Fig2}(a) shows a nearly monochromatic response at $\nu_0 \approx 2$ Hz. Correspondingly, the Fourier transform of $n_2$ in Fig.~\ref{Fig2}(c) shows peaks at even multiples of $\nu_0$, with $2\nu_0$ being the most dominant one. The factor of 2 arises because $g(y)$ decreases both for positive and negative $y$, see Eq.~\eqref{alpha2}. The peak at zero frequency appears because the atom-cavity coupling leads to the scattering of photons into cavity mode 2 even when the particle is stationary. We also note that the pump frequency, $\omega_p$, is $\sim$ $2\pi\times$ 384 THz (the system Hamiltonian in Eq.~(\ref{H_RF}) is written in the rotating frame of the pump field) and we are looking for a small modulation of a few Hz ($\sim 4$ Hz) on top of this frequency. Such a modulation can be measured by a heterodyne measurement of a portion of the on-axis pump laser interfered with the cavity output field. This measurement needs to be done with a laser of very narrow line-width (less than 1 Hz) which is within the reach of the available technologies. The output signal at 2$\nu_0\approx 4$ Hz corresponds to a temperature of $\sim 0.2$ nK, which is quite low. We can use a weakly outcoupled  atom laser to generate a very low temperature atomic beam of atoms such that on an average only one atom passes through the cavity at a time. Alternatively, we can use a BEC from which single atoms with known trajectories are extracted using interfering (Bragg) laser beams. We also observe a non-linear time evolution of $x_a$ as revealed via a series of frequency peaks that are integer multiples of a fundamental frequency in the Fourier transform of $e^{2ikx_a}$ which overlaps exactly with the Fourier transform of $e^{i\phi_{2,ss}}$ (see Fig.~\ref{Fig2}(d)). Thus, the cavity field $\alpha_2$ can be used to reconstruct the snake state trajectories; see Appendix \ref{recons} for full reconstruction. In Appendix \ref{n1phi1}, we show the time evolution of $n_1$ and $\phi_1$ and observe that they also predominantly oscillate at frequency $2\nu_0$ with the caveat that the oscillation amplitude normalized by the corresponding time-average value is much smaller.
With a heterodyne measurement and external locking of the laser frequency (e.g. to a reference vapor cell), the drift of the laser frequency can be minimized and the laser linewidth can be reduced. With the available experimental techniques, the laser intensity can be stabilised to 1$\%$ level or below. We also note from Fig.~\ref{Fig2}(b) that the change in the number of photons in the cavity during the snake state evolution is close to $100\%$ which makes it easier to detect.

Additionally, we would like to point out that the effect of Doppler shift on the spontaneous emission from an atom moving in a linear single mode Fabry-P\'erot cavity and its effect on the resulting transmission spectrum \cite{Meystre1992} is also a relevant issue since the atoms are moving in a complex snake state trajectory. A full treatment of this is somewhat beyond the scope of the current manuscript, and it is also complicated by the fact that we consider a ring cavity structure where an emitted photon by one cavity mode is absorbed by another cavity mode. Nevertheless, we have included a brief analysis of this issue in Appendix \ref{appenDShift} by generalizing the treatment given in \cite{Kozlovskii2001}.

\begin{figure}
\centering
\includegraphics[width=1\columnwidth, height=0.9\columnwidth]{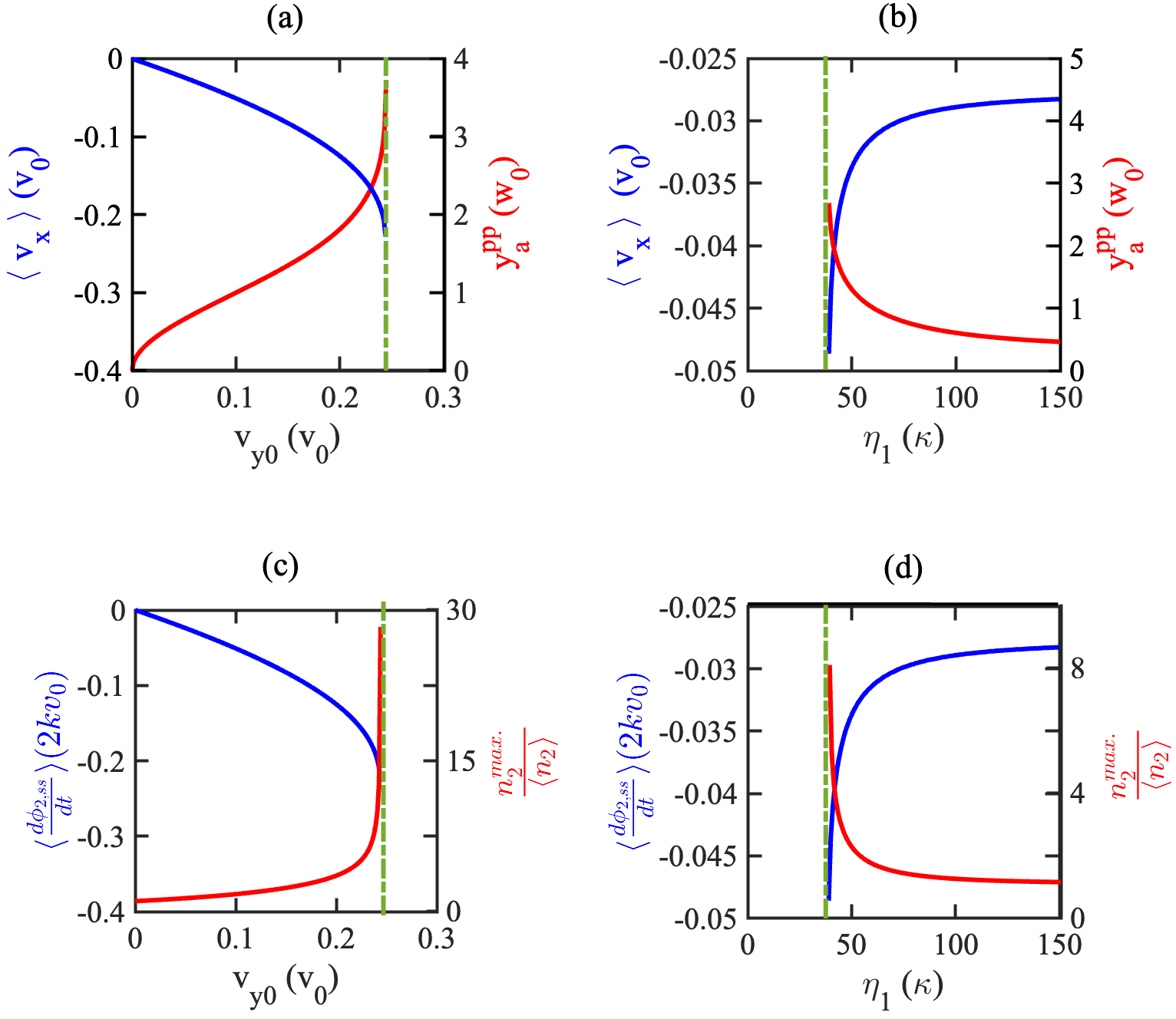}
\caption{\textit{(color online)}: Average drift velocity, $\langle v_x\rangle$ ($blue$ curve), and peak-to-peak $y_a$ variation, $y^{pp}_a$ ($red$ curve),
as a function of (a) the initial velocity $v_{y0}$ (with $\eta_1=80\kappa$ and $\eta_2=0$) and (b) the pump strength $\eta_1$  (with $v_{y0}=0.06v_0$ and $\eta_2=0$).
Here $v_0$ is a natural velocity scale of the system (see text), $w_0$ is the cavity mode waist and
the initial speed along $x$-direction is fixed to 0.
The normalized maximum photon number in cavity mode 2, $n_2^{max.}/\langle n_2\rangle$ ($red$ curve), and the average time derivative of the corresponding phase, $\langle d \phi_{2,ss}/dt\rangle$ ($blue$ curve),
are shown in (c) as a function
of $v_{y0}$ and in (d) as a function of $\eta_1$. The vertical dashed-dotted $green$ lines mark the boundary where
snake state trajectory ceases to exist, and the particle is not trapped along the $y$-direction.}\label{Fig3}
\end{figure}

\section{Manipulation of snake states}\label{manipulation}
Next, we discuss how the snake state trajectories can be manipulated from the perspective of one-dimensional transport along $x$-direction. We quantify such a transport by two properties: average drift velocity, $\langle v_x\rangle$ which is proportional to the particle conductivity along the $x$-direction and peak-to-peak amplitude $y^{pp}_{a}$ of $y_a$ which signifies the deviation from a purely one-dimensional transport. We choose two tuning parameters: The initial speed $v_{y0}$ of the atom and the pump strength $\eta_1$. As shown in Fig.~\ref{Fig3}(a),  we observe an increase in $\langle v_x\rangle$ ($blue$ curve) accompanied by an increase in $y^{pp}_{a}$ ($red$ curve) when $v_{y0}$ is increased. On the other hand, both these trends are reversed when we instead increase $\eta_1$ as depicted in Fig.~\ref{Fig3}(b). Such a behaviour can be understood from a basic Lorentz force picture, noting that $|B(y)| \propto \eta^2_{1}$. Above (below) a critical $v_{y0}$ ($\eta_1$), the atom cannot be trapped along the $y$-direction by the synthetic magnetic field, and the snake state trajectory picture breaks down. This boundary is marked by the vertical dashed-dotted $green$ lines in Fig.~\ref{Fig3}. The initial velocity of the atom, $v_{y0}$, can be controlled either by trapping the atom in an optical tweezer and then spatially accelerating the tweezer or by using a thermal atomic beam which has a specific distribution of speeds.

\begin{figure}
\centering
\includegraphics[width=.9\columnwidth, height=.5\columnwidth]{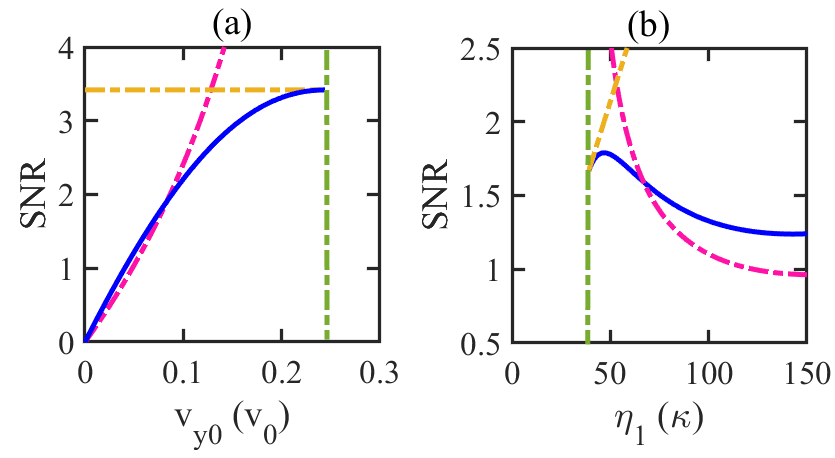}
\caption{\textit{(color online)}: Signal-to-noise ratio as a function of (a) the initial velocity $v_{y0}$ (with $\eta_1=80\kappa$ and $\eta_2=0$) and (b) the pump strength $\eta_1$ (with $v_{y0}=0.06v_0$ and $\eta_2=0$). The $blue$ curve shows the numerically obtained SNR. The $yellow$ dashed-dotted curve represents the analytical approximation of the SNR for $y_a^{pp}>>w_0$ and $magenta$ dashed-dotted curve represents the analytical approximation of the SNR for $y_a^{pp}<<w_0$ as discussed in Section \ref{manipulation}. The $green$ dashed-dotted lines represent the boundaries where snake state trajectory picture breaks down.}\label{FigSNR}
\end{figure}

Fig.~\ref{Fig3}(c, d) shows the maximum photon number variation $ n_2^{max.}$ in cavity mode 2 normalized by $\langle n_2\rangle$ and the average time variation of the corresponding phase $\langle d \phi_{2,ss} /dt\rangle$ as a function of $v_{y0}$ and $\eta_1$. We observe that these two quantities mimic the behaviour of $y^{pp}_{a}$ and $\langle v_x\rangle$ and thus, the snake state trajectories can be mapped on the cavity field for a wide-range of parameters.. Beyond the critical points where the particle is not trapped, the photon number $n_2$ decays to zero as the particle leaves the cavity mode. In Appendix \ref{freq_vel}, we provide simple scalings of various quantities as a function of our tuning parameters and give some examples of the time evolution of particle properties and cavity fields for different initial conditions. Finally, we note that we have assumed that the initial position of the particle $y_{a0}=0$ in all the examples presented here. For $|y_{a0}|\approx w_0$, we find out that the particle performs normal ($cyclotron$) orbits similar to those in a homogeneous magnetic field which can also be mapped on the cavity field of mode 2; see Appendix \ref{phase_diag} for details.

Next, we look at the feasibility of detection of number of photons leaking from the cavity from the shot noise point of view. We look at the signal-to-noise ratio (SNR) to distinguish the maximum ($n_2^{max.}$) and the minimum ($n^{min.}_2$) photon number and define SNR as $S/N$ where the signal S is 
$
S=n_2^{max.}-n^{min.}_2 .
$
The shot noise associated with this signal is $N=\sqrt{n_2^{max.}+n^{min.}_2}$. We plot SNR as a function of the initial velocity, $v_{y0}$ in Fig.~\ref{FigSNR}(a) and as a function of the pump strength, $\eta_1$ in Fig.~\ref{FigSNR}(b). The key feature which we observe is that SNR is high for high $y_a^{pp}$ but for low $y_a^{pp}$, the SNR can be below 1.
%We define the minimum detectable $y$-variation of the snake trajectory $\Delta y_a^{det.}$ for which the corresponding signal to noise ratio for the photon number detection is 1. We plot the minimum detectable amplitude of the snake state trajectory in Fig.~\ref{FigSNR}(c, d) as a function of the initial velocity, $v_{y0}$ and the pump strength, $\eta_1$ and observe that it is easier to detect snake states with large $y_a^{pp}$.
To understand how the SNR can be improved, we use Eq.~(\ref{alpha2}) to approximate SNR in two limits. For $y_a^{pp} << w_0$, we can perform Taylor expansion around $y_a = 0$ and obtain
\bea
\text{SNR} 
&=&\sqrt{\frac{n_2^{max.}}{2}}\left(\frac{y_a^{pp}}{w_0}\right)^2
\eea 
for distinguishing the maximum and the minimum photon number and which is plotted in $magenta$ in Fig.~\ref{FigSNR}(a,b) and agrees with the numerically obtained SNR in the small amplitude regimes which appear at small initial velocities and large pump strengths. For $y_a^{pp} >> w_0$, $n^{min.}_2 \approx 0$ and thus SNR $=\sqrt{n_2^{max.}}$ which is plotted in Fig.~\ref{FigSNR}(a,b) in $yellow$ color and agrees with the numerically obtained SNR for large amplitude regimes. 
%Similarly, we can also obtain the minimum detectable amplitude for $y_a^{pp} << w_0$ and obtain
%\[
%\Delta y_a^{det} = \frac{w_0}{2}\left(\frac{2}{n_2^{max.}}\right)^{1/4}.
%\]
%This result is shown in Fig.~\ref{FigSNR}(c,d) in $purple$. 
From these results, we see that the SNR can be improved by increasing $n_2^{max.}$ while keeping $y_a^{pp}$ constant which can be done by tuning other parameters as shown in Eq.~(\ref{alpha2}).

\begin{figure}
\centering
\includegraphics[width=1\columnwidth, height=1.55\columnwidth]{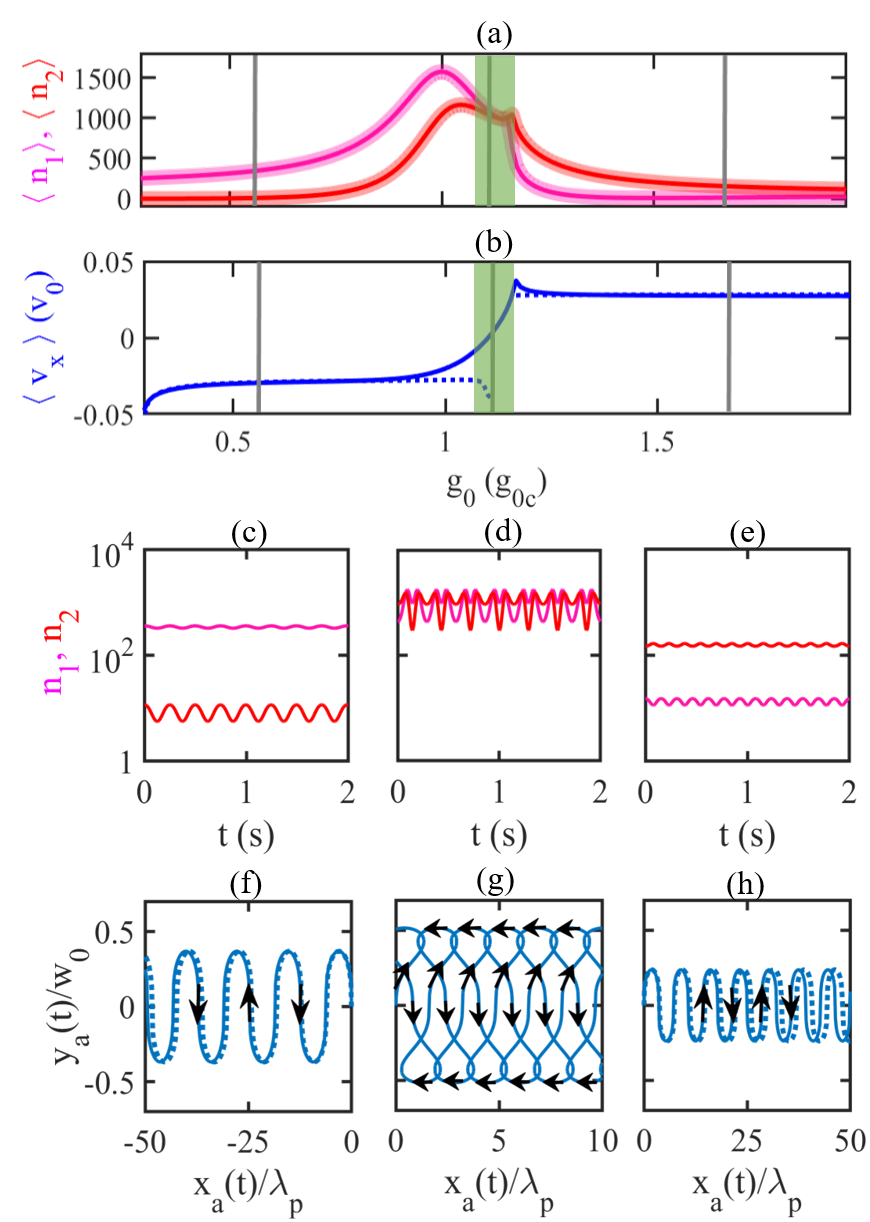}
\caption{\textit{(color online)}: (a) Average photon number variation in cavity mode 1 ($magenta$ curve) and 2 ($red$ curve) as a function of $g_0$. The faded region denotes the corresponding peak-to-peak amplitude around the average photon number. The vertical $grey$ lines correspond to $g_0 = (0.55, 1.1, 1.6) g_{0c}$ and represent the parameters used for plotting (c, f), (d, g), and (e, h), respectively. (b) Average drift velocity, $\langle v_x\rangle$ is plotted as a function of the coupling strength $g_0$ which is normalized by $g_{0c}=2\pi\times 90$ MHz, see text. The dotted $blue$ curve shows the average drift speed in the absence of cavity feedback. The $green$ shaded region in (a, b) marks the region where the snake state trajectory is destroyed due to dynamical cavity feedback. (c-e) The evolution of the photon number in cavity mode 1, $n_1$ ($magenta$ curves), and mode 2, $n_2$ ($red$ curves), with time $t$ and (f-h) the corresponding particle trajectories in the $x$-$y$ plane for $g_0 = (0.55, 1.1, 1.6) g_{0c}$, respectively. The $dotted$ trajectories in (f-h) depict the particle trajectories without cavity back-action and the $black$ arrows indicate the direction of increasing time in the time evolution. For all the calculations in this figure, $\eta_1=80\kappa$, $\eta_2=0$, $v_{x0}=0$ and $v_{y0}=0.06v_0$ is considered, which gives $g_{0}=0.29g_{0c}$ as the minimum coupling parameter required to get a trapped atomic trajectory.}\label{Fig4}
\end{figure}

\section{Cavity feedback induced breakdown of snake state trajectories}\label{breakdown}
We now discuss two different effects on the snake state trajectories which arise from strong atom-cavity coupling $g_0$. From Eq.~(\ref{alpha12}) and (\ref{alpha2}), we note that $\alpha_1 \propto \eta_1$ and $\alpha_2 \propto \eta_1g^2_0$ (for $\eta_2=0$) and this distinction allows us to tune the relative number of photons in the two cavity modes. This is illustrated in Fig.~\ref{Fig4}(a, c, e) where $\langle n_{1}\rangle$ is much larger(smaller) than $\langle n_{2}\rangle$ for $g_0$ much smaller(larger) than $g_{0c}$. Here $g_{0c}\simeq \sqrt{\vert\Delta_a\vert \vert\Delta_c\vert/2}$ is obtained by setting $\vert\alpha_1\vert=\vert\alpha_2\vert$ in Eq.~(\ref{alpha12}) and assuming $y_a=0=x_a$. Such a control on the sign of $n_1-n_2$ allows us to tune the sign of the induced magnetic field gradient and thus the directionality of the generated snake state trajectories, see Fig.~\ref{Fig4}(f, h) (same starting conditions: $x_{a0}=0=y_{a0}$ for $t=0$ are used in both the cases). In Fig.~\ref{Fig4}(b), we show the corresponding behaviour of $\langle v_x\rangle$ which is negative(positive) for $g_0<<(>>)g_{0c}$.

The situation around $g_0 \approx g_{0c}$ ($green$ shaded region in Fig.~\ref{Fig4}(a, b)) is more complicated, where the dynamical cavity feedback leads to the breakdown of the snake state trajectories. In this regime, the particle is still trapped near the cavity axis, but the resultant trajectory has a different topology as compared to the snake states; see Fig.~\ref{Fig4}(g) for an example of such a trajectory. The origin of such complicated trajectories can be understood by looking at the corresponding time evolution of $n_{1}$ and $n_{2}$, see Fig.~\ref{Fig4}(d) where we note that the sign of ($n_1-n_2$) and hence the generated magnetic field gradient changes its amplitude and sign with time. The faded region around the $\langle n_{1(2)}\rangle$ curves in Fig.~\ref{Fig4}(a) shows the peak-to-peak deviations from the mean value, and we find that the breakdown regime overlaps well with the region where the faded regions of $\langle n_1\rangle$ and $\langle n_2\rangle$ overlap (see Fig.~\ref{Fig4}(d)). To illustrate that such a breakdown happens due to cavity feedback, we have additionally plotted $\langle v_x\rangle$ ($dotted$ line in Fig.~\ref{Fig4}(b)) and atomic trajectories ($dotted$ trajectories in Fig.~\ref{Fig4}(f, h) and in Fig.~\ref{Fig5Appen} in
Appendix \ref{breakdown_trajec}) for the case without feedback by fixing the photon number to the time-averaged values of the feedback case. For $\vert\langle n_1\rangle-\langle n_2\rangle\vert\gg 0$, the resultant snake state trajectories in the two cases are very comparable (see discussion related to Fig. \ref{Fig1}(d) as well). For $\vert\langle n_1\rangle-\langle n_2\rangle\vert\approx 0$, the atom is not trapped, which appears as a gap in the $dotted $ curve in Fig.~\ref{Fig4}(b). We finally note that a similar breakdown of snake state trajectories can be achieved by pumping both the cavity modes such that $\eta_1 \approx \eta_2$, which leads to comparable photon numbers in the two cavity modes.

\section{Conclusions and outlook}\label{conclusions}
This work focuses on realizing the atomic analogue of electronic snake state trajectories in a ring-cavity coupled to a single two-level atom. We have shown that atom-cavity interaction in such a set-up creates an effective spatially varying magnetic field with its strength depending on the difference in the photon number in the two counter-propagating running wave cavity modes, which cannot be achieved in a standing-wave cavity. Atom in such a non-uniform perpendicular magnetic field follows snake state trajectories and can be detected by monitoring the output cavity fields as they dynamically depend on the atom's position. The atomic snake state trajectories provide an advantage over their electronic counterparts found in condensed matter systems, where the charge carriers interact strongly with the system making their manipulation difficult. The cold-atom surroundings allow us to engineer the properties of atomic snake states by changing the system parameters, such as the initial velocity of the atom, external pump strength, and atom-cavity coupling strength. We can also tune the effect of cavity back-action via atom-cavity coupling strength to change the topology of snake states and create even richer dynamics. 

Our proposed set-up and methodology can be straightforwardly extended to induce magnetic fields with more intricate spatial structure by using multi-mode cavities \cite{Ballantine} and can be used to detect the resulting topological trajectories via the output cavity fields. As a further extension of this work, one can study the behaviour of a Bose-Einstein condensate in the presence of such a gauge field where the interplay of atom-atom interactions, atom-cavity interaction and cavity feedback can give rise to exotic topological phases of matter \cite{Qi2011,Seo2012,Sato2009} and non-linear instabilities \cite{Wu2002,Ferris2008}. It will also be exciting to study how such dynamical magnetic field-induced snake states compete with other well known phenomena in high finesse ring cavities like superradiant Rayleigh scattering and collective recoil lasing \cite{RitschRev, RitschRev2,Courteille}. Finally, we would also like to point out that determining whether such snake states have distinct topological features like the edge states in conventional quantum Hall systems in solid-state devices requires a lattice-based calculation, which is not within the scope of the current manuscript and may be carried out in the future.

\section{Acknowledgements}
We thank Manuele Landini, Farokh Mivehvar, B. Prasanna Venkatesh, Mishkatul Bhattacharya and Puja Mondal,
for a number of helpful discussions at various
stages of this work. SG thanks the ETH group of T.
Esslinger and P. \"Ohberg at Heriot-Watt University for
helpful discussions during his visits there. This work is supported by a BRNS (DAE, Govt. of India)
Grant No. 21/07/2015-BRNS/35041 (DAE SRC Outstanding Investigator scheme). PS was also supported by a UGC (Government of India) fellowship at the initial stage of this work.

\appendix
\setcounter{figure}{0} \renewcommand{\thefigure}{A.\arabic{figure}}
\counterwithin{figure}{section}
\section{Derivation of the system Hamiltonian}\label{SH}
In this section, we derive the system Hamiltonian given in Eq.~(\ref{H_RF}) of the main text. The single particle Hamiltonian describing the coupled atom-cavity system is $\hat{H}_{SP}=\hat{H}_{A}+\hat{H}_{C}+\hat{H}_{A-C}$, where the atomic and cavity part of the Hamiltonian are respectively given as -
\bea
\hat{H}_{A} &=& \frac{\vec{P}^2}{2m_a}\hat{\mathbb{I}}+\frac{\hbar \omega_a \hat{\sigma}_z}{2}  \\
\hat{H}_C & = & \hbar \omega_{c} (\hat{a}_1^{\dagger}\hat{a}_{1}+  \hat{a}_2^{\dagger}\hat{a}_{2} )+\hbar \eta_1 (\hat{a}_1 e^{i\omega_p t} + \hat{a}_1^{\dag} e^{-i\omega_p t})\nn\\
&+& \hbar \eta_2 (\hat{a}_2 e^{i\omega_p t} + \hat{a}_2^{\dag} e^{-i\omega_p t})
\eea 
Here $E_e - E_g = \hbar\omega_a$. The single two-level excited atom scatters the photons into the two cavity modes.
The atom-cavity interaction is given as -
\begin{eqnarray}
\hat{H}_{int.}&=& \hat{H}_{A-C}= -\vec{d}\cdot\vec{E}_{C} \label{Hint.}
\end{eqnarray}
where $\vec{d}=d(\hat{\sigma}^{+}+\hat{\sigma}^{-})$ is the dipole operator with $\hat{\sigma}^{+}=\vert e\rangle\langle g\vert$, and, $\hat{\sigma}^{-}=\vert g\rangle\langle e\vert$. 
$\hat{H}_{A-C}$ describes the interaction between the atom and the cavity fields (polarized along $y$ and $z$ directions) in one arm of the ring cavity, and its corresponding electric field is given by -
\bea
\vec{E}_{C}(\vec{r})&=&\hat{e}_y \sqrt{\frac{\hbar \omega_c}{2\epsilon_0 V}}e^{-y^2/w^2_0}\left(\hat{a}_1 e^{ikx} +\hat{a}^{\dag}_{1} e^{-ikx}\right)\nn\\
&+& \hat{e}_z \sqrt{\frac{\hbar \omega_c}{2\epsilon_0 V}}e^{-y^2/w^2_0}\left(\hat{a}_2 e^{-ikx} +\hat{a}^{\dag}_{2} e^{ikx}\right) \label{fields} 
\eea
Here $\epsilon_0$ is the vacuum permittivity and $V$ is the mode volume.
To get a clearer picture, we move to the interaction picture. The atomic field operators are given as $\hat{\sigma}^{\pm}(t) =  \hat{\sigma}^{\pm}(0) e^{\pm i\omega_a t}$. The time evolution of the cavity field operators is written as $\hat{a}_{1(2)}(t) =  \hat{a}_{1(2)}(0) e^{-i\omega_c t}$. Similarly, for $\hat{a}^{\dag}_{1,2}$, we get - $\hat{a}_{1,2}^{\dag}(t) = \hat{a}_{1,2}^{\dag}(0) e^{i\omega_c t}$. 
Using (\ref{fields}), the atom-cavity field interaction in the interaction picture takes the following form -
\begin{eqnarray}
\hat{H}^{I}_{A-C}&=&-\vec{d}\cdot\vec{E}_{C}\nn\\
&=& \hbar g_1(y)\Big[\hat{\sigma}^{+}(t)\hat{a}_1(t)e^{ikx}+\hat{\sigma}^{-}(t)\hat{a}_1(t)e^{ikx}\nn\\
&+&\hat{\sigma}^{+}(t)\hat{a}^{\dag}_1(t)e^{-ikx}+\hat{\sigma}^{-}(t)\hat{a}_1^{\dag}(t)e^{ikx}\Big]\nn\\
&+& \hbar g_2(y)\Big[\hat{\sigma}^{+}(t)\hat{a}_2(t)e^{-ikx}+\hat{\sigma}^{-}(t)\hat{a}_2(t)e^{-ikx}\nn\\
&+&\hat{\sigma}^{+}(t)\hat{a}^{\dag}_2(t)e^{ikx}+\hat{\sigma}^{-}(t)\hat{a}^{\dag}_2(t)e^{ikx}\Big]
\end{eqnarray} 
If $\omega_a\sim\omega_c$, then the terms with $e^{\pm i (\omega_a-\omega_c)t}$ will have small transition amplitudes that are proportional to $\frac{1}{(\omega_a+\omega_c)^2}$.
Therefore, the fast oscillating terms with frequency $\omega_a+\omega_c$ can be neglected as compared to the slow oscillating terms with frequency $\omega_a-\omega_c$.  Transforming back to the Schr\"odinger picture we get \cite{Shore,Sakurai, Nicacio} -
\begin{eqnarray}
\begin{aligned}
\hat{H}_{A-C} &=& \hbar g_1(y) \left[\hat{\sigma}^{+}\hat{a}_1 e^{ikx}+\hat{\sigma}^{-}\hat{a}_1^{\dag}e^{-ikx}\right]\nn\\ 
&+& \hbar g_2(y) \left[\hat{\sigma}^{+}\hat{a}_2e^{-ikx}+\hat{\sigma}^{-}\hat{a}^{\dag}_2e^{ikx}\right]
\end{aligned}
\end{eqnarray}
The transformation to the rotating frame of the pump field is carried through the unitary operator, 
\begin{equation*}
\hat{U}(t)=e^{-i\omega_p t (\frac{\hat{\sigma}_z}{2} + \hat{a}_1^{\dag}\hat{a}_1 +\hat{a}_2^{\dag}\hat{a}_2)}
\end{equation*}
Since the observables and the states respectively transform as -
$\hat{O}_{RF} = \hat{U}^{\dag} \hat{O} \hat{U}$, and, 
$\vert \Psi_{RF} \rangle = \hat{U}^{\dag} \vert \Psi\rangle$
in the rotating frame, the Schr\"odinger equation transforms as -
\begin{align*}
i\hbar \frac{\partial}{\partial t}\vert\Psi_{RF}\rangle &= i\hbar \left[\frac{\partial}{\partial t}\left(\hat{U}^{\dag}\vert\Psi\rangle\right)\right]= \hat{H}_{RF}  \vert \Psi_{RF}\rangle
\end{align*}
where 
\begin{equation}
\hat{H}_{RF} = \frac{-\hbar \omega_p \hat{\sigma}_z}{2} -\hbar \omega_p \hat{a}^{\dag}_{1}\hat{a}_{1}-\hbar \omega_p \hat{a}^{\dag}_{2}\hat{a}_{2} + \hat{U}^{\dag} \hat{H}_{SP} \hat{U}
\end{equation}
is the single-particle Hamiltonian in the rotating frame of the pump field. To get $\hat{U}^{\dag}\hat{H}_{SP}\hat{U}$, we use the Baker-Hausdorff formula which finally gives us
\begin{eqnarray}
\hat{H}_{RF} &=& \frac{\hat{P}^2}{2m_a}\hat{\mathbb{I}}-\frac{\hbar \Delta_a \hat{\sigma}_z}{2} -\hbar\Delta_c\left(\hat{a}^{\dag}_{1}\hat{a}_{1}+\hat{a}^{\dag}_{2}\hat{a}_2\right)\nn\\
&+&\hbar\eta_1\left(\hat{a}^{\dag}_1+\hat{a}_1\right)+\hbar\eta_2\left(\hat{a}^{\dag}_2+\hat{a}_2\right)\nn\\
&+&\hbar g_1(y) 		\left[\hat{\sigma}^{+}\hat{a}_1 e^{ikx}+\hat{\sigma}^{-}\hat{a}_1^{\dag}e^{-ikx}\right]\nn\\ 
&+& \hbar g_2(y) \left[\hat{\sigma}^{+}\hat{a}_2e^{-ikx}+\hat{\sigma}^{-}\hat{a}^{\dag}_2e^{ikx}\right]
\end{eqnarray}
where $\Delta_a = \omega_p -\omega_a$ is the atom-pump detuning and $\Delta_c = \omega_p-\omega_c$ is the cavity-pump detuning. This is the system Hamiltonian considered in Eq.~(\ref{H_RF}) of the main text.

The interaction Hamiltonian, $\hat{H}_{I}$ in the bare-state basis contains off-diagonal terms and can be written as -
\begin{eqnarray}
\hat{H}_I&=& \begin{bmatrix}
-\frac{\hbar}{2}\Delta_a + C_t  & \hbar c_1\\
\hbar c^{*}_1& \frac{\hbar}{2}\Delta_a +C_t\nn\\
\end{bmatrix}
\end{eqnarray}
where we have used \\
$C_t = -\hbar \Delta_c (\vert \alpha_1\vert^2 + \vert \alpha_2\vert^2 ) + 2\hbar \eta_1 \vert \alpha_1\vert \cos\phi_1 + 2\hbar \eta_2 \vert \alpha_2\vert \cos\phi_2 $, and
$c_1 =  g(y)\alpha_1 e^{ikx}+ g(y)\alpha_2e^{-ikx},$

We diagonalise the Hamiltonian $\hat{H}_{I}$ in the space spanned by the atom-photon bare-states, namely $\vert e,\alpha_1,\alpha_2\rangle, $ and $\vert g,\alpha_1,\alpha_2\rangle$, and obtain the following eigenstates referred as dressed states $\vert D_1\rangle$ and $\vert D_2\rangle$, for the coupled atom-photon system, with $E_{1}$ and $E_{2}$ as their eigenvalues, respectively. They are 
\begin{subequations}
\begin{eqnarray}
E_1 &=&-\hbar \Delta_c (\vert \alpha_1\vert^2 + \vert \alpha_2\vert^2)+  2\hbar \eta_1 \vert \alpha_1\vert \cos\phi_1\nn\\
&+& 2\hbar \eta_2 \vert \alpha_2\vert \cos\phi_2 - \frac{\hbar G}{2};\label{E1}
\\
\vert D_1\rangle &=& \frac{1}{\sqrt{2G(G+\Delta_a)}}\begin{bmatrix}
G+\Delta_a\\
-2c^{*}_{1}
\end{bmatrix}\label{D1_a}
\\
E_2 &=&-\hbar \Delta_c (\vert \alpha_1\vert^2 + \vert \alpha_2\vert^2)+  2\hbar \eta_1 \vert \alpha_1\vert \cos\phi_1 \nn\\
&+& 2\hbar \eta_2 \vert \alpha_2\vert \cos\phi_2 + \frac{\hbar G}{2};\label{E2}
\\
\vert D_2\rangle &=& \frac{1}{\sqrt{2G(G+\Delta_a)}}\begin{bmatrix}
2c_1\\
G+\Delta_a
\end{bmatrix}\label{D2_a}
\end{eqnarray}
\end{subequations}
where
$
\vert c_1\vert^2 = \left(g^2(y)\vert \alpha_1\vert^2 +g^2(y)\vert \alpha_2\vert^2 \right)$ and
$G =\sqrt{\Delta_a^2 + 4\vert c_1\vert^2}\nn
$

Now, we derive the equation of motion for the probability amplitude, $\psi_1$, to find
the atom in the lowest energy dressed state, $\vert D_1\rangle$. In the dressed state basis ($\ket{D_j}$ basis) for the internal Hilbert space of the atom at any point $\vec{r}$, the full state vector of the atom and the corresponding equation of motion is \cite{Dalibard} -
\beq
\vert \Psi (\vec{r},t)\rangle = \sum_{j=1,2} \psi_j(\vec{r},t)\vert D_{j}\rangle
\eeq
\beq 
i \hbar \frac{\partial}{\partial t}\vert \Psi (\vec{r},t)\rangle = \hat{H}_{RF}\vert \Psi (\vec{r},t)\rangle \label{Heqn} 
\eeq
The action of the momentum operator, $\hat{P}$, on the atomic wavefunction, $\vert \Psi(\vec{r},t)\rangle$ is given as \cite{Dalibard} -
\begin{eqnarray}
\hat{P}\vert \Psi(\vec{r},t)\rangle &=& -i\hbar \vec{\nabla}\left[\sum_{j=1,2} \psi_j(\vec{r},t)\vert D_{j}\rangle\right]\nn\\ 
&=& -i\hbar \sum_{j}\left[(\vec{\nabla}\psi_j(\vec{r},t)) \vert D_{j}\rangle + \psi_j(\vec{r},t) (\vec{\nabla}\vert D_{j}\rangle)\right]\nn\\
&=& \sum_{j,l=1,2} \left[\vec{p}\delta_{l,j} - \vec{\textbf{A}}_{l,j}\right]\psi_j \vert D_{l}\rangle
\end{eqnarray}  
where $\vec{\textbf{A}}_{l,j}=i\hbar \langle D_{l}\vert\vec{\nabla}\vert D_{j} \rangle $ is the vector potential and $\vec{p}=-i\hbar \vec{\nabla}$ does not act on the spinorial part.
From this, we can straightforwardly write the kinetic energy term as 
\begin {widetext}
\beq 
\frac{\vec{P}^2}{2m_a}\vert \Psi (\vec{r},t)\rangle=\frac{1}{2m_a}\sum_{j,l,m=1,2}\Big\lbrace\left(\vec{p}\delta_{l,j} - \vec{\textbf{A}}_{l,j}\right)\left[\left(\vec{p}\delta_{m,l} - \vec{\textbf{A}}_{m,l}\right)\psi_j \right]\Big\rbrace\vert D_{m}\rangle
\eeq 
\end{widetext} 
We can write down a $2\times 2$ matrix $\vec{\textbf{A}}$ whose components are given as - 
$ \vec{\textbf{A}}_{l,j}=i\hbar\langle D_l\vert\vec{\nabla}\vert D_j\rangle $.
We project the Schr\"odinger equation (\ref{Heqn}) to the lowest energy dressed state, $\vert D_1\rangle$, to obtain Eq.~(\ref{EOM}) of the main text.

\section{}
\subsection{Effect of the scalar potential, $W$, on the magnetic field, $\vec{B}$}\label{appen_scalar}
\begin{figure}
\centering
\includegraphics[width=.75\columnwidth, height=.6\columnwidth]{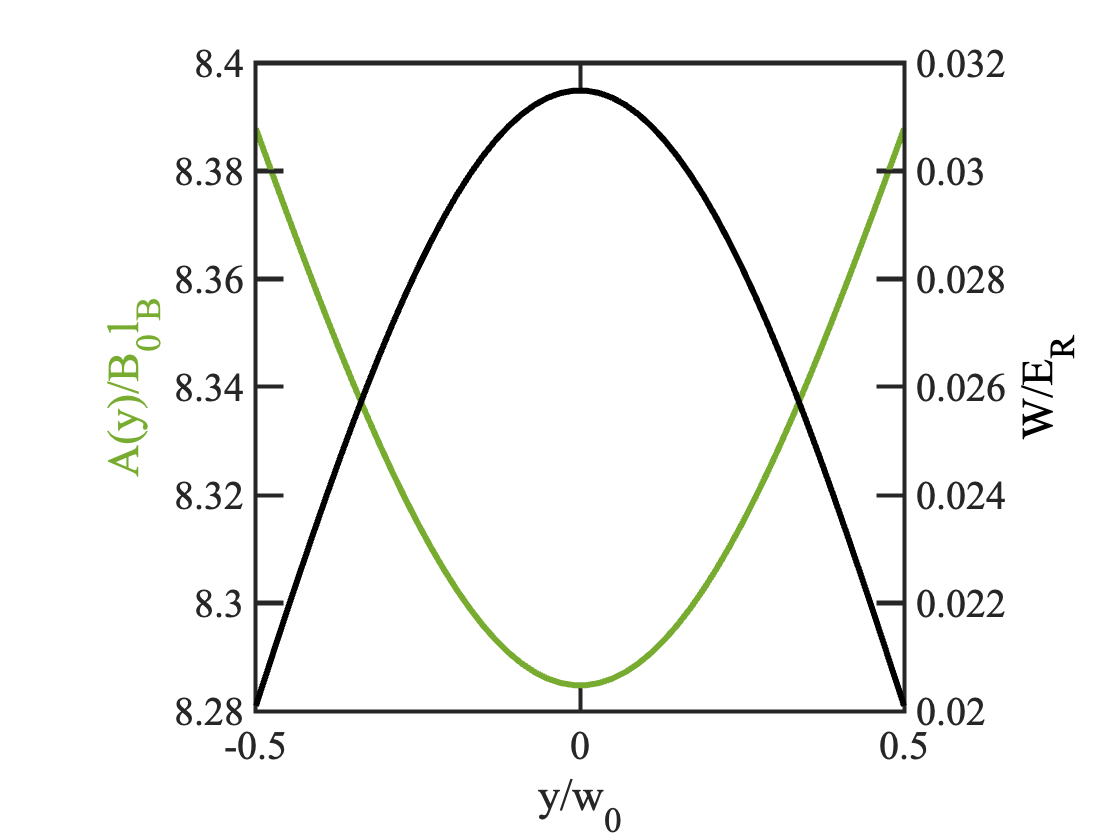}\label{Figvecsca}
\caption{\textit{(color online)}: Comparison of scalar potential \textit{W} and vector potential \textit{A}.}\label{Fig_vecsca}
\end{figure}

We provide a comparison of the vector and the scalar potentials in Fig.~\ref{Fig_vecsca} which shows that the scalar potential has negligible contribution to the system dynamics. We can switch to a multi-level description which allows for the possibility of magic wavelengths where the effect of the scalar potential can be cancelled. 
\subsection{Effect of an additional trapping potential on the magnetic field, $\vec{B}$}\label{appen_exttrap}
%We require a two-dimensional harmonic trap to observe a snake trajectory. If we include such a trap with frequency, $\omega_0$, in the considered system, we get the following modified Hamiltonian -
%\begin{eqnarray}
%\hat{H}_{RF}=\hat{H}_0+\hat{H}_{I}
%\end{eqnarray}
%where 
%\begin{eqnarray}
%\hat{H}_0 & = & \frac{\hat{P}^2}{2m_a}\hat{\mathbb{I}}\nn \\
%\hat{H}_{I} &=& -\frac{\hbar \Delta_a\hat{\sigma}_z}{2} -\hbar\Delta_c\left(\hat{a}^{\dag}_{1}\hat{a}_{1}+\hat{a}^{\dag}_{2}\hat{a}_2\right)\nn\\
%&+&\hbar\eta_1\left(\hat{a}_1 + \hat{a}^{\dag}_{1}\right) + \hbar\eta_2\left(\hat{a}_2 + \hat{a}^{\dag}_{2}\right)\nn\\
%&+& \hbar  \left(g_1(y)\hat{\sigma}^{+}\hat{a}_1 e^{ikx} +g_2(y)\hat{\sigma}^{+}\hat{a}_2e^{-ikx}  \right.\nn \\
%&+& \left. g_1(y)\hat{\sigma}^{-}\hat{a}^{\dag}_1e^{-ikx}+g_2(y)\hat{\sigma}^{-}\hat{a}^{\dag}_2e^{ikx} \right)\nn\\
%&+& \frac{1}{2}m_a\omega_0^2(x^2+y^2)\hat{\mathbb{I}}\nn
%\end{eqnarray}
The off-diagonal terms of the interaction Hamiltonian, $\hat{H}_I$, which couple the bare states of the atom-cavity system are mainly responsible for the shape and emergence of a magnetic field. An external harmonic trapping potential provides only an additional confining potential apart from the dressed state energy, $E_1$, in Eq. (2) of the main text.
For a harmonic trap with a frequency of $\Omega_0\approx 2\pi\times50$ Hz, the typical length scales are approximately 4 $\mu m$. However, for the parameters considered in this work, the length scales associated with snake trajectories are of the order of $\approx 9$ $\mu m$ and hence would make the detection of the snake states difficult. Therefore, instead of a harmonic trap we can use a free atomic beam which is not trapped in the $x-y$ plane and thus does not restrict the length scale of the trajectory, making their detection possible.

\section{}
\subsection{Reconstruction of atomic trajectory from output cavity fields }\label{recons}
\begin{figure}
\centering
\includegraphics[width=0.8\columnwidth, height=0.8\columnwidth]{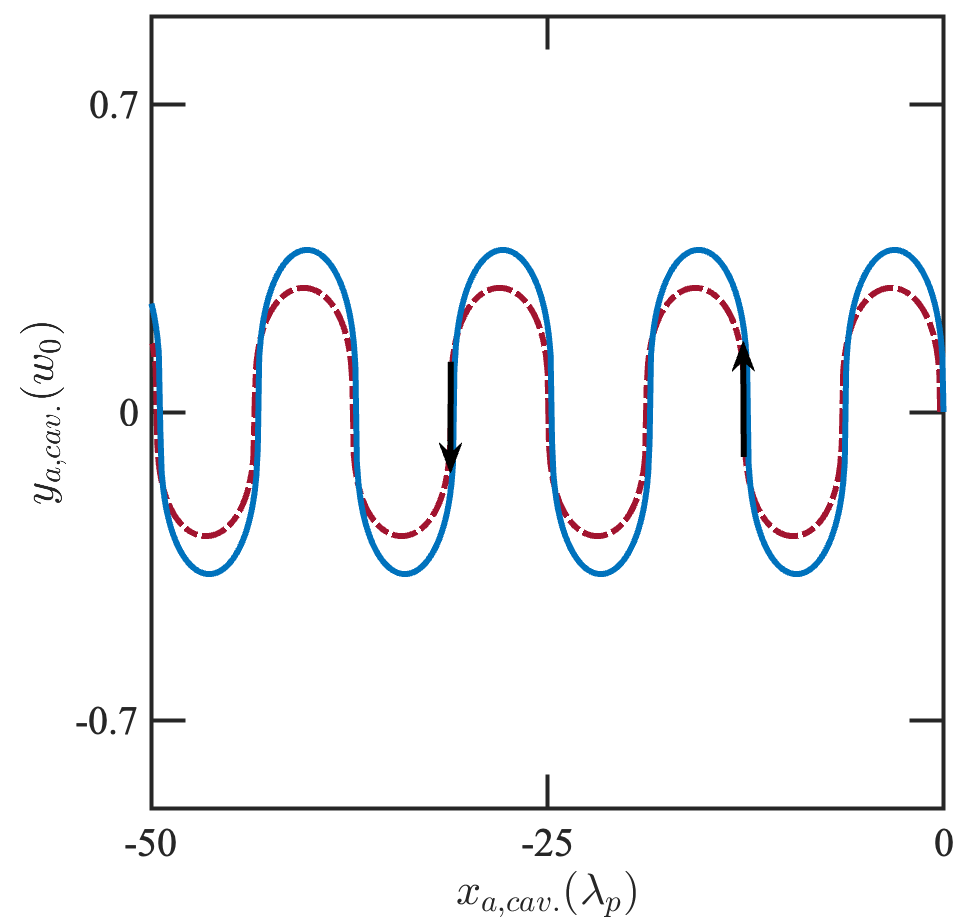}
\caption{\textit{(color online)}: \textit{Solid line} shows the snake state trajectory and the \textit{dotted line} shows the reconstructed snake state trajectory using parameters of the output cavity field $\alpha_2$. Refer text for details.}\label{Fig1Appen}
\end{figure}

We use the output from cavity mode 2 to recreate the snake state trajectory of the atom using the approximations considered in Section \ref{Cavity_detection}. The $x$-position of the atom can be estimated by the phase $\phi_{2,ss}$ given as $\frac{x_{a,cav.}}{\lambda_p} = \frac{\phi_{2,ss}}{2k\lambda_p}$ and the $y$- position of the atom can be estimated from the values of $n_2$ given by the approximated formula: 
\begin{equation}
\frac{y_{a,cav.}}{w_0}=\frac{1}{2}\sqrt{log_{10}\left(\frac{n_2^{max.}}{n_2}\right)},\label{yacav}
\end{equation}
where we have inverted the formula for $\vert\alpha_2\vert$ given in Eq.~(\ref{alpha2}) and have used $\frac{\eta_1^2g_0^4}{\Delta_a^2(\bar{\Delta}_c^2+\bar{\kappa}^2)^2}\simeq n_2^{max.}$ (keeping $y=0$). Experimentally, we can measure $n_2^{max.}$ directly, so, the $y-$ reconstruction will be closer to the actual trajectory. With these formulae, we reconstruct the snake state trajectory in Fig.~\ref{Fig1Appen}. We also invert the sign of reconstructed $y_a$ when it reaches zero because $n_2$ does not carry this information explicitly. A variation in the peak-to-peak values of $y_{a,cav}$ arises due to the approximations considered. 
%\textcolor{magenta}{The velocity of the atom along the direction of propagation ($x-$ direction in this case) is small as compared to the speed of the emitted photons and the natural velocity scale of the system defined by $v_0=\hbar/(m_al_B)$ (see Fig. \ref{Fig_vx}, this is the adiabatic approximation mentioned in Section II of the main text) and this allows us to neglect the Doppler effect.
%\begin{figure}
%\centering
%\includegraphics[width=.75\columnwidth, height=.65\columnwidth]{vx.png}
%\caption{\textit{(color online)}: \textcolor{magenta}{Velocity of the atom along the ($x-$) propagation direction.}}\label{Fig_vx}
%\end{figure}}

\subsection{Photon numbers in mode 1 and their phase $\phi_1$}\label{n1phi1}
\begin{figure}[H]
\centering
\includegraphics[width=1\columnwidth, height=1\columnwidth]{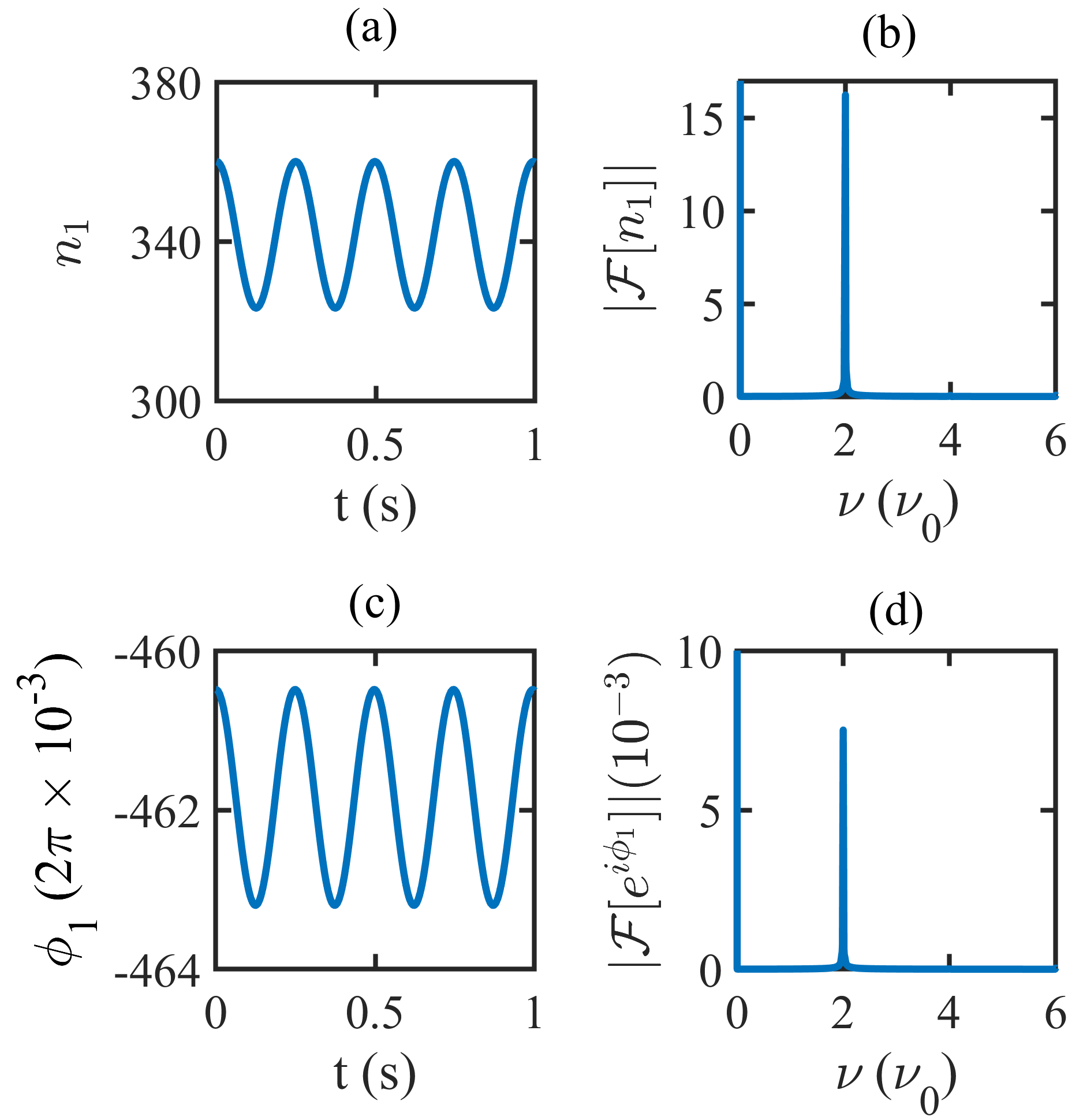}
\caption{\textit{(color online)}: (a) Photon numbers in cavity mode 1, $n_1$, as a function of time, $t$. (b) Fourier transform of $n_1$ which shows a peak at $2\nu_0$. (c) Phase $\phi_1$ of photons in mode 1 as a function of time, $t$. (d) Fourier transform of $e^{i\phi_1}$ which also shows a peak at $2\nu_0$. }\label{Fig2Appen}
\end{figure}

We provide the plots for the photon numbers in mode 1 and its Fourier transform, which shows a peak at $2\nu_0$ in Fig.~\ref{Fig2Appen}(a, b). The peaks at $4\nu_0$ and $6\nu_0$  are not visible here, but their amplitude increases for higher initial speeds along the $y$-direction. Phase $\phi_1$ of photons in mode 1 (see Fig.~\ref{Fig2Appen}(c)) shows a small oscillation amplitude which arises from the $y_a$ dependence of $U$ and $\gamma$. The Fourier transform of $ e^{i\phi_1}$ shows a small amplitude peak at $2\nu_0$ (see Fig.~\ref{Fig2Appen}(d)) which is different from the Fourier transform of $e^{i\phi_{2,ss}}$ as there is no $x_a$ dependence in $\phi_1$.

\subsection{Effect of the Doppler shift on the atomic trajectory}\label{appenDShift}
The motion of the atom with respect to the propagation direction of the cavity field could give rise to a Doppler shift. The Doppler shift is given by $\vec{k}\cdot\vec{v}$ where $\vec{k}$ is the wavevector of the cavity field ($\vec{k}=k\hat{x}$ for cavity mode $\hat{a}_1$ and $\vec{k}=-k\hat{x}$ is for cavity mode $\hat{a}_2$) and $\vec{v}$ is the velocity of the atom. 
%We include the Doppler shift in the system Hamiltonian (\ref{H_RF}) of the main text to get
%$
%\hat{H}_{RF} = \hat{H}_{0}+\hat{H}_{I},
%$ 
%where 
%\bea 
%\hat{H}_0 & = & \frac{\hat{P}^2}{2m_a}\hat{\mathbb{I}}\nn \\
%\hat{H}_{I} &=& -\frac{\hbar \Delta_a\hat{\sigma}_z}{2} -\hbar\Delta_c\left(\hat{a}^{\dag}_{1}\hat{a}_{1}+\hat{a}^{\dag}_{2}\hat{a}_2\right)\nn\\
%&+&\hbar\eta_1\left(\hat{a}_1 + \hat{a}^{\dag}_{1}\right) + \hbar\eta_2\left(\hat{a}_2 + \hat{a}^{\dag}_{2}\right)\nn\\
%&+& \hbar  \left(g_1(y)\hat{\sigma}^{+}\hat{a}_1 e^{i(kx+\vec{k}\cdot\vec{v}t)} +g_2(y)\hat{\sigma}^{+}\hat{a}_2e^{-i(kx-\vec{k}\cdot\vec{v}t)}  \right.\nn \\
%& +  & \left. g_1(y)\hat{\sigma}^{-}\hat{a}^{\dag}_1e^{-i(kx+\vec{k}\cdot\vec{v}t)}+g_2(y)\hat{\sigma}^{-}\hat{a}^{\dag}_2e^{i(kx-\vec{k}\cdot\vec{v}t)} \right).\nn\\
%\eea 
\begin{figure}
\centering
\includegraphics[width=0.8\columnwidth, height=0.8 \columnwidth] {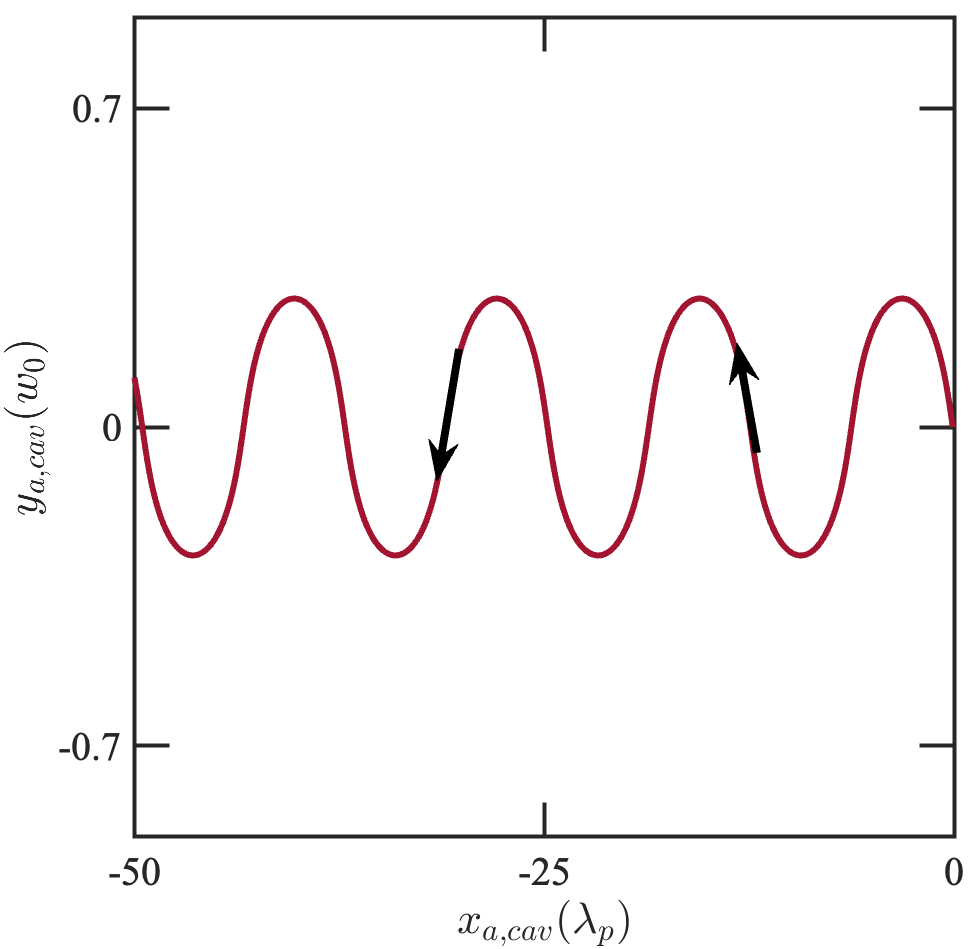}
\caption{({\it color online}) The reconstructed snake state trajectory including the effect of the Doppler shift.}\label{Figrecons_DS}
\end{figure}
\begin{figure}
\centering
\includegraphics[width=0.8\columnwidth, height=0.8 \columnwidth] {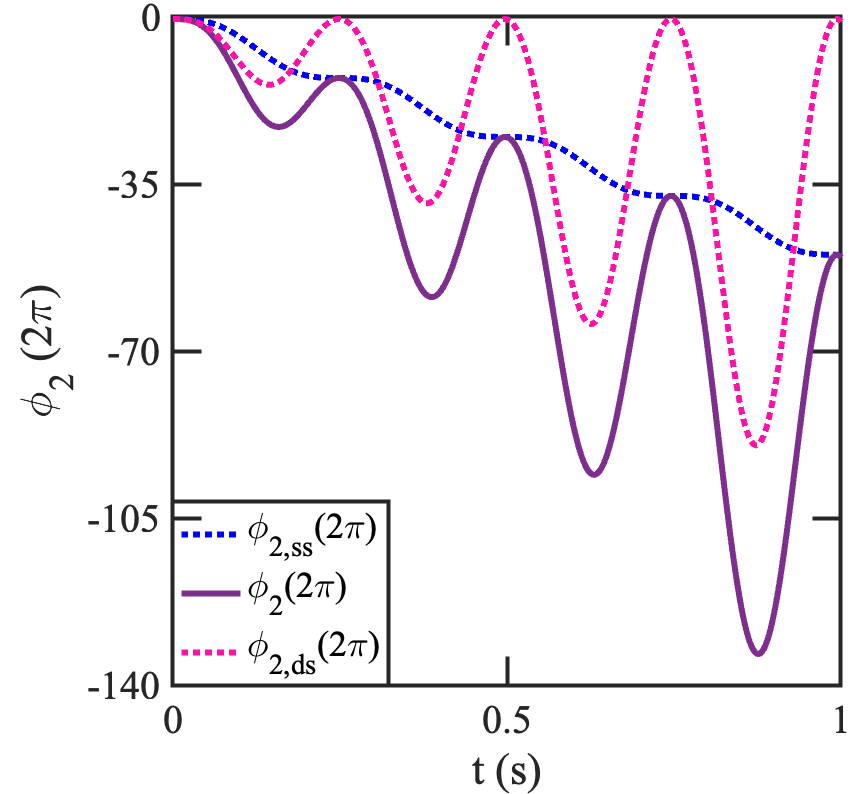}
\caption{({\it color online}) Full phase, $\phi_2$, of cavity mode 2 as a function of time $t$ showing the contribution of $\phi_{2,ss}$ and $\phi_{2,ds}$ including the effect of the Doppler shift.}\label{FigDShift}
\end{figure}
The modified steady state expression for the two cavity fields after including the effect of the Doppler shift is given as
\begin{eqnarray}
\alpha_{1(2)} &=&\langle \hat{a}_{1(2)} \rangle = \langle D_1 \vert \hat{a}_{1(2)} \vert D_1\rangle\nn\\
&=& \frac{i\eta_{1(2)}(i\bar{\Delta}_{c}-\bar{\kappa}) + i\eta_{2(1)}(iU+\gamma)e^{\mp 2ikx_a(t)+2i\vec{k}\cdot\vec{v}t}}{(i\bar{\Delta}_{c}-\bar{\kappa})^2-(iU+\gamma)^2} \nn\\ \label{alphaDShift}
\end{eqnarray}
Solving Eq.~(\ref{eom_vx_f}, \ref{eom_vy_f}) simultaneously with Eq.~ (\ref{alphaDShift}) (assuming similar initial conditions used in Section \ref{trajectories} of the main text), we observe in Fig.~\ref{Figrecons_DS} that the resulting snake state trajectories do not show any effect of the Doppler shift. This is because the Doppler shifts for the two counter-propagating fields are opposite and hence have no effect on the resulting atomic trajectories. However, the time variation of phase $\phi_2$ is now given as
\begin{equation}
\phi_2\approx \phi_{2,ss}+\phi_{2,ds}=2kx_a(t)+2\vec{k}\cdot \vec{v}t
\end{equation}
and is shown in Fig.~\ref{FigDShift}. An increase in the oscillation amplitude of $\phi_2$ arises due to the time dependence of the Doppler shift component ($\phi_{2,ds}\approx 2\vec{k}\cdot \vec{v}t$) in the phase.
The wavevector for cavity mode 2 is $-k\hat{x}$ and the atom is moving along $-x$ direction which allows us to write 
\begin{eqnarray}
\phi_2 &=& 2kx_a(t)+2kt\frac{dx_a}{dt}\nn \\ 
\Rightarrow \frac{dx_a}{dt}&=&-\frac{x_a}{t}+\frac{\phi_2}{2kt}
\end{eqnarray}
We solve the above differential equation numerically to obtain $x_{a,cav}$, and use $y_{a,cav}$ from Eq.~(\ref{yacav}) to reconstruct the snake state trajectory of the atom in Fig.~\ref{Figrecons_DS} from the cavity 2 output. A variation in the peak-to-peak values of $\textit{y}_{a,cav}$ arises due to the approximations considered.

\section{}
\subsection{Snake state trajectories for different initial conditions}\label{freq_vel}

\begin{figure}
\centering
\includegraphics[width=1\columnwidth, height=1.8\columnwidth]{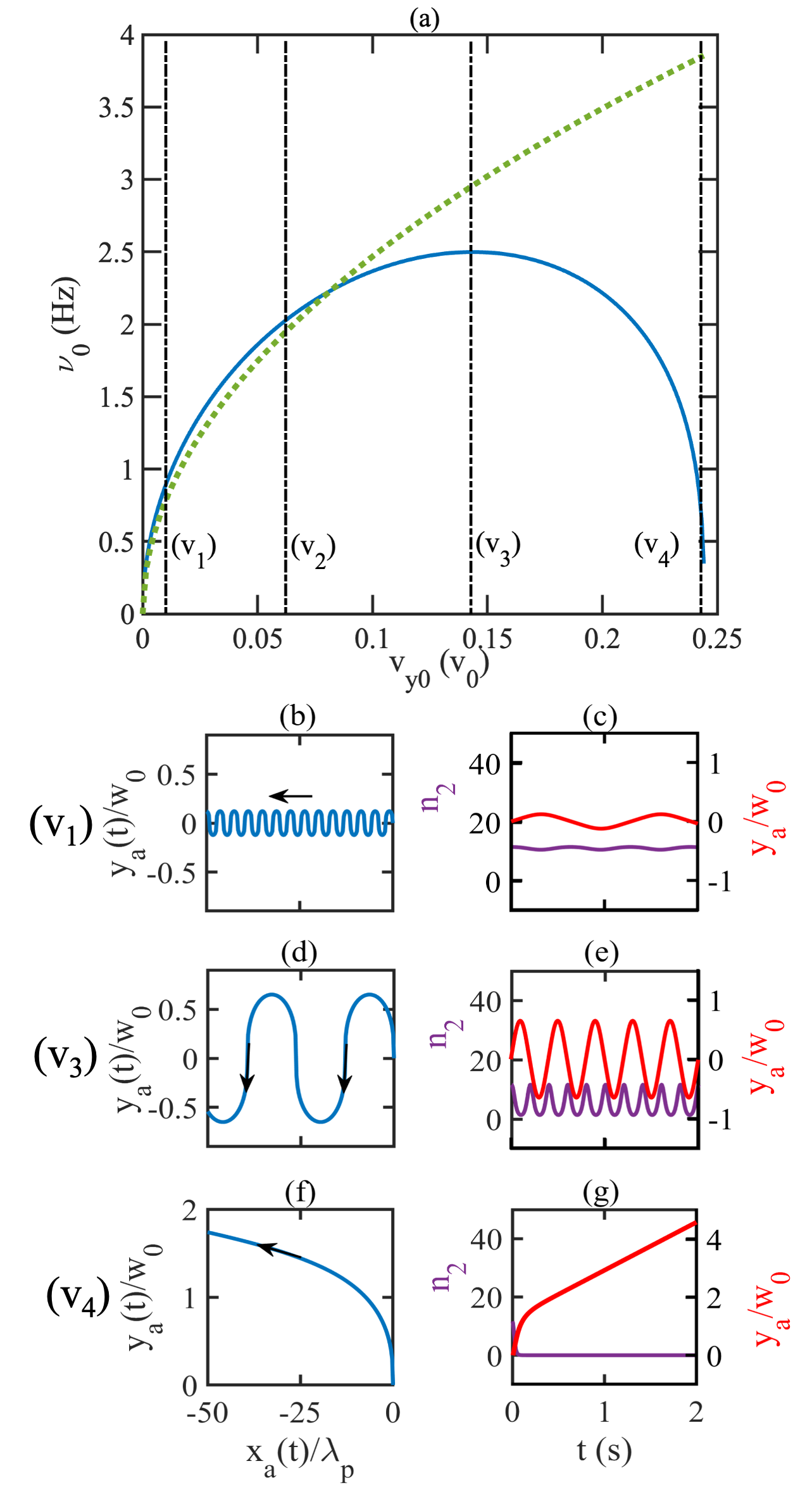}
\caption{\textit{(color online)}: (a) Oscillation frequency $\nu_0$ of atomic position $y_a$ along $y$-direction is plotted as a function of initial velocity, $v_{y0}$ along the $y$-direction for fixed $v_{x0}=0$. The $green$ $dotted$ line shows the analytical frequency obtained from the $y_{period}$ formula in Eq.~(\ref{yperiod}). We have considered four points $(v_1)$, $(v_2)$, $(v_3)$ and $(v_4)$ of initial velocity and show the corresponding trajectories and photon numbers. The $v_{y0}$ velocity at point $(v_2)$ has been considered in the main text of the paper and therefore, its trajectories are not shown here. Snake state trajectories for $v_{y0}=v_1$, $v_3$ and $v_4$ are shown in (b), (d) and (f), respectively. Correspondingly, photon number ($purple$ curve) in mode 2 ($left$ $y$-axis) and $y_a/w_0$ ($red$ curve) variation ($right$ $y$-axis) as a function of time are shown in (c), (e) and (g). For parameters and explanation, refer text. }\label{Fig3Appen}
\end{figure}
We look at the structure of snake state trajectories for different initial speeds $v_{y0}$ along the $y$-direction. We consider $\eta_1 = 80\kappa, \eta_2 = 0, x_0=0=y_0$ and $v_{x0}=0$. We plot the peak frequency $\nu_0$ of the Fourier transform of $y_a$ in Fig.~\ref{Fig3Appen}(a) and observe two different regimes. Up to $v_{y0}\approx 0.13v_0$, $\nu_0$ increases. In this regime, the oscillation amplitude along $y$-direction increases with increasing $v_{y0}$ due to the increased initial speed and the amplitude of $n_2$ increases correspondingly as shown in  Fig.~\ref{Fig3Appen}(c, e). The spatial period along the $x$-direction also increases, as shown in Fig.~\ref{Fig3Appen}(b, d). The $green$ $dotted$ curve in Fig.~\ref{Fig3Appen}(a) also shows that the analytical frequency obtained from the $y_{period}$ formula in Eq.~(\ref{yperiod}) fits better for small initial velocities, $v_{y0}$.

For initial velocities higher than $0.13v_0$, $\nu_0$ decreases, and the amplitude of the snake state trajectories along the $y$-direction increases. For $v_{y0} > 0.24v_0$, the atom cannot be trapped along the $y$-direction by the synthetic magnetic field (see Fig.~\ref{Fig3Appen}(f)) and the corresponding photon number $n_2$ decays to zero, see Fig.~\ref{Fig3Appen}(g).
\begin{figure}[H]
\centering
\includegraphics[width=0.9\columnwidth, height=0.9\columnwidth]{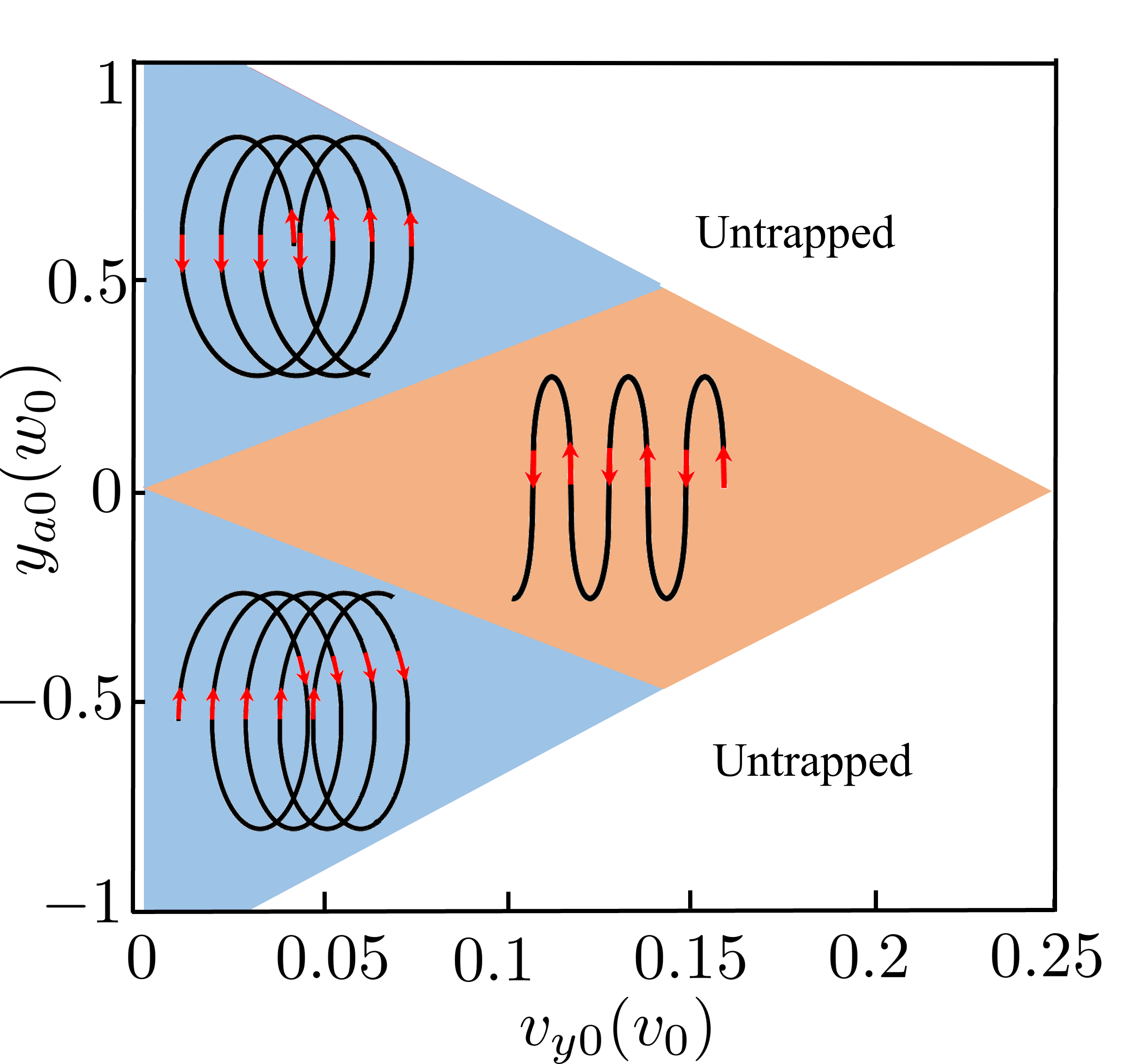}
\caption{\textit{(color online)}: Phase diagram for the atomic trajectories - in $blue$ shaded regions, the particle follows $cyclotron$ orbits, in $orange$ shaded region, the particle follows snake state trajectories, and the $white$ region indicates the regime where the atom is not trapped by the synthetic magnetic field. $y_{a0}$ and $v_{y0}$ are the initial position and speed of the atom along the $y$-direction, respectively. $w_0$ is the waist of the cavity mode and $v_0$ is a natural scale of the particle speed, see text. The $red$ arrows indicate the direction of particle evolution with increasing time.}\label{Fig4Appen}
\end{figure}

\subsection{Phase diagram of the atomic trajectories}\label{phase_diag}
The atom subjected to the perpendicular non-uniform artificial magnetic field can follow snake state trajectories or normal orbits depending on the initial velocity and initial position along the $y$- axis. We provide a qualitative phase diagram to visualize the various regimes in Fig.~\ref{Fig4Appen}  for $g_0 = 0.55g_{0c}$, $\eta_1 = 80\kappa$ and $\eta_2=0$. When the atom is far away ($\vert y_a\vert > 0.5w_0$) from the cavity centre, it sees a nearly uniform magnetic field with a small  slope (see Fig.~\ref{Fig1}(b) of the main text) and performs $cyclotron$ orbits ($blue$ shaded region) for small $v_{y0}$ values. For high $v_{y0}$ values, the uniform magnetic field is not strong enough to keep the atom trapped, and therefore, it escapes the cavity. We obtain clockwise(anti-clockwise) normal orbits for $y_{a0}<(>)0$. When the atom starts near the cavity centre ($x_{a0}=0$ and $|y_a|<0.5w_0$), it follows normal orbits for very small $v_{y0}$. An increase in $v_{y0}$ increases the range of $y_{a0}$ where the atom follows snake state trajectories ($orange$ shaded region) for $v_{y0}>0$ as it sees a spatially varying magnetic field which reverses its sign along $y=0$ axis. For $0.13v_0<v_{y0}<0.24v_0$, the range of $y_{a0}$, where the snake state motion is allowed, decreases. The atom escapes the cavity for $v_{y0}>0.24v_0$. Changing $\eta_1$ gives a similar phase diagram with an increase in the value of escape velocity for increasing $\eta_1$. 

\subsection{Atomic trajectories in the breakdown regime}\label{breakdown_trajec}
\begin{figure}[H]
\centering
\includegraphics[width=1\columnwidth, height=0.6\columnwidth]{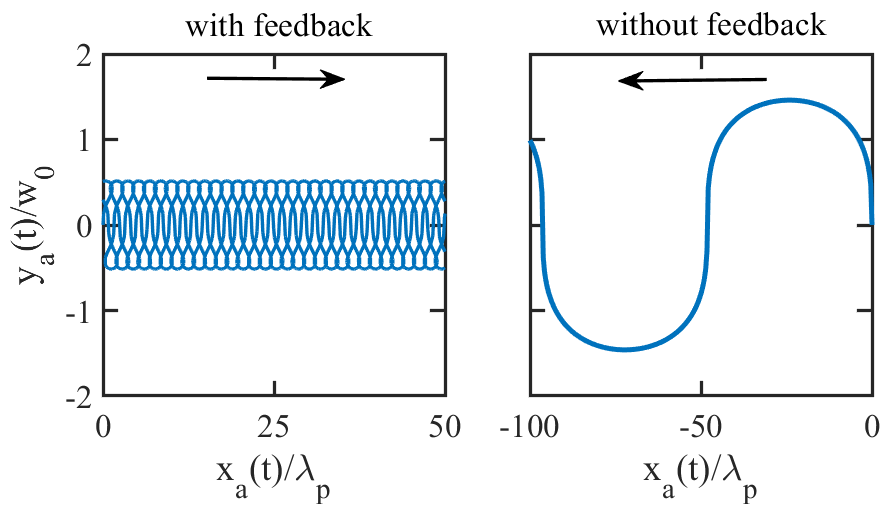}
\caption{\textit{(color online)}: For $g_{0}=1.1 g_{0c}$, we show the trajectories of the atom with and without feedback ($\langle n_1\rangle =946$ and $\langle n_2\rangle=894$). The initial velocity $v_{x0}=0$ and $v_{y0}=0.06v_0$.}\label{Fig5Appen}
\end{figure}

We plot the atomic trajectories in the breakdown regime for the coupling strength $g_0=1.1 g_{0c}$ with and without considering the cavity feedback in Fig.~\ref{Fig5Appen}. The absence of feedback results in a left-moving snake-like trajectory as opposed to the case when we consider the effect of cavity feedback on the atomic trajectory, which results in a complicated right-moving trajectory. 


\begin{thebibliography}{60}
\def \bb{\bibitem}

%\bb{Madison} \textbf{Vortex Formation in a Stirred Bose-Einstein Condensate
%}, K. W. Madison, F. Chevy, W. Wohlleben and J. Dalibard, \href{https://journals.aps.org/prl/pdf/10.1103/PhysRevLett.84.806}{Phys. Rev. Lett. \textbf{84}, 806 (2000)}.

%\bb{Wasim}\textbf{Quantum gas microscopy for single atom and spin detection}, C. Gross and W. S. Bakr, \href{https://www.nature.com/articles/s41567-021-01370-5#citeas}{Nat. Phys. \textbf{17}, 1316(2021).}
%
%\bb{Kaufman}\textbf{Quantum science with optical tweezer arrays of ultracold atoms and molecules}, A. M. Kaufman and K.-K. Ni, \href{https://www.nature.com/articles/s41567-021-01357-2}{Nat. Phys. \textbf{17}, 1324(2021).}

\bb{Dalibard} \textbf{Colloquium: Artificial gauge potentials for neutral atoms}, J. Dalibard, F. Gerbier, G. Juzeli\~unas and P. \"Ohberg, \href{https://journals.aps.org/rmp/pdf/10.1103/RevModPhys.83.1523}{Rev. Mod. Phys., \textbf{83}, 1523 (2011).}

\bb{Goldman} \textbf{Light-induced gauge fields for ultracold atoms}, N. Goldman, G. Juzeli\~unas, P. \"Ohberg, and I. B. Spielman, \href{https://iopscience.iop.org/article/10.1088/0034-4885/77/12/126401/pdf}
{Rep. Prog. Phys. \textbf{77}, 126401 (2014).}

\bb{Spielman}\textbf{Raman processes and effective gauge potentials}, I. B. Spielman, \href{https://journals.aps.org/pra/abstract/10.1103/PhysRevA.79.063613}{ Phys. Rev. A \textbf{79},063613 (2009).}

\bb{Gerbier} \textbf{Gauge fields for ultracold atoms in optical superlattices}, F. Gerbier and J. Dalibard, \href{https://iopscience.iop.org/article/10.1088/1367-2630/12/3/033007}{New J. Phys., \textbf{12}(3), 033007 (2010).}

\bb{Jaksch} \textbf{Creation of effective magnetic fields in optical lattices: the Hofstadter butterfly for cold neutral atoms}, D. Jaksch and P. Zoller, \href{https://iopscience.iop.org/article/10.1088/1367-2630/5/1/356}{New J. Phys. \textbf{5}, 56 (2003).}

\bb{Sorensen} \textbf{Fractional Quantum Hall States of Atoms in Optical Lattices}, A. S. Sorensen, E. Demler, and M. D. Lukin, \href{https://journals.aps.org/prl/pdf/10.1103/PhysRevLett.94.086803}{Phys. Rev.
Lett. \textbf{94}, 086803 (2005).} 

\bb{Kolovsky} \textbf{Creating artificial magnetic fields for cold atoms by photon-assisted tunneling}, A. R. Kolovsky, \href{https://iopscience.iop.org/article/10.1209/0295-5075/93/20003}{Europhys. Lett. \textbf{93}, 20003 (2011).}

\bb{Cooper} \textbf{Optical Flux Lattices for Ultracold Atomic Gases}, N. R. Cooper, \href{https://physics.aps.org/featured-article-pdf/10.1103/PhysRevLett.106.175301}{Phys. Rev. Lett. \textbf{106}, 175301 (2011).} 

\bb{JStruck}\textbf{Quantum Simulation of Frustrated Classical Magnetism in Triangular Optical Lattices}, J. Struck, C. \"Olschl\"ager, R. Le Targat, P. Soltan-Panahi, A. Eckardt, M. Lewenstein, P. Windpassinger, and K.
Sengstock, \href{https://www.science.org/doi/10.1126/science.1207239}{Science \textbf{333}, 996 (2011).}

\bb{MAidels}\textbf{Experimental Realization of Strong Effective Magnetic Fields in an Optical Lattice}, {M. Aidelsburger, M. Atala, S. Nascimbène, S. Trotzky, Y.-A.
Chen, and I. Bloch}, \href{https://journals.aps.org/prl/abstract/10.1103/PhysRevLett.107.255301}{Phys. Rev. Lett. \textbf{107}, 255301 (2011).}

\bb{PHauke} \textbf{Non-Abelian Gauge Fields and Topological Insulators in Shaken Optical Lattices}, P. Hauke, O. Tieleman, A. Celi, C. \"Olschl\"ager, J. Simonet,
J. Struck, M. Weinberg, P. Windpassinger, K. Sengstock,
M. Lewenstein, and A. Eckardt, \href{https://journals.aps.org/prl/abstract/10.1103/PhysRevLett.109.145301}{Phys. Rev. Lett. \textbf{109},
145301 (2012).}

\bb{JStruckPHauke} \textbf{Tunable Gauge Potential for Neutral and Spinless Particles in Driven Optical Lattices}, J. Struck, C. \"Olschl\"ager, M. Weinberg, P. Hauke, J. Simonet,
A. Eckardt, M. Lewenstein, K. Sengstock, and P.
Windpassinger, \href{https://journals.aps.org/prl/abstract/10.1103/PhysRevLett.108.225304}{Phys. Rev. Lett. \textbf{108}, 225304 (2012).}

\bb{Miyake}\textbf{Realizing the Harper Hamiltonian with Laser-Assisted Tunneling in Optical Lattices}, H. Miyake, G. A. Siviloglou, C. J. Kennedy, W. C. Burton,
and W. Ketterle, \href{https://journals.aps.org/prl/pdf/10.1103/PhysRevLett.111.185302}{Phys. Rev. Lett. \textbf{111}, 185302 (2013).}

\bb{MAtala}\textbf{Realization of the Hofstadter Hamiltonian with Ultracold Atoms in Optical Lattices}, M. Aidelsburger, M. Atala, M. Lohse, J. T. Barreiro, B.
Paredes, and I. Bloch, \href{https://journals.aps.org/prl/abstract/10.1103/PhysRevLett.111.185301}{Phys. Rev. Lett. \textbf{111}, 185301 (2013).}

\bb{Lin_spielman}\textbf{Bose-Einstein Condensate in a Uniform Light-Induced Vector Potential}, Y.-J. Lin, R. L. Compton, A. R. Perry, W. D. Phillips, J. V.
Porto, and I. B. Spielman, \href{https://journals.aps.org/prl/abstract/10.1103/PhysRevLett.102.130401}{Phys. Rev. Lett. \textbf{102}, 130401
(2009).}

\bb{LinCompton}\textbf{Synthetic magnetic fields for ultracold neutral atoms}, Y.-J. Lin, R. L. Compton, K. Jimenez-Garcia, J. V. Porto,
and I. B. Spielman, \href{https://www.nature.com/articles/nature08609}{Nature (London) \textbf{462}, 628 (2009).}

\bb{Lin_Garcia}\textbf{Spin–orbit-coupled Bose–Einstein condensates}
Y.-J. Lin, K. Jim\'enez-Garc\'ia and I. B. Spielman, \href{https://www.nature.com/articles/nature09887}{Nature \textbf{471}, 83 (2011).}

\bb{Feder}\textbf{Synthetic spin-orbit interactions
and magnetic fields in ring-cavity QED}, F. Mivehvar and D. L. Feder, \href{https://journals.aps.org/pra/abstract/10.1103/PhysRevA.89.013803}{Phys. Rev. A \textbf{89}, 013803 (2014)}.

\bb{Bloch} \textbf{Many-body physics with ultracold gases}, I. Bloch, J. Dalibard, and W. Zwerger, \href{https://journals.aps.org/rmp/pdf/10.1103/RevModPhys.80.885}{Rev. Mod. Phys. \textbf{80}, 885 (2008).}

\bb{Buluta} \textbf{Quantum Simulators}, I. Buluta and F. Nori, \href{https://science.sciencemag.org/content/sci/326/5949/108.full.pdf}{Science \textbf{326}, 108 (2009).}

\bb{Blochrev}\textbf{Ultra cold quantum gases in optical lattices}, I. Bloch, \href{https://www.nature.com/articles/nphys138.pdf}
{Nature Phys. \textbf{1}, 23 (2005).}

\bb{Schmeidmayer} \textbf{Quantum wires and quantum dots for neutral atoms}, J. Schmeidmayer, \href{https://epjd.epj.org/articles/epjd/abs/1998/10/d8076/d8076.html}{Eur. Phys. J. D \textbf{4}, 57 (1998).}

\bb{Cooper2008} \textbf{Rapidly rotating atomic gases}, N. R. Cooper, \href{https://doi.org/10.1080/00018730802564122}{Adv.
in Phys. \textbf{57}, 539 (2008).}

\bb{Viefers} \textbf{Quantum Hall physics in rotating Bose-Einstein
condensates}, S. Viefers, \href{https://iopscience.iop.org/article/10.1088/0953-8984/20/12/123202}{J. Phys. Cond. Matt. \textbf{20},
123202 (2008).}

\bb{Fetter}\textbf{Rotating trapped Bose-Einstein condensates}, A.
L. Fetter, \href{https://journals.aps.org/rmp/abstract/10.1103/RevModPhys.81.647}{Rev. Mod. Phys. \textbf{81}, 647 (2009).}

\bb{Ruhman}\textbf{Topological States in a
One-Dimensional Fermi Gas with Attractive Interaction}, J. Ruhman, E. Berg, and E. Altman,
\href{https://journals.aps.org/prl/abstract/10.1103/PhysRevLett.114.100401}{Phys. Rev. Lett. 114, 100401 (2015).}

\bb{Jiang}\textbf{Majorana Fermions in Equilibrium and in
Driven Cold-Atom Quantum Wires}, L. Jiang, T. Kitagawa, J. Alicea, A. R. Akhmerov, D. Pekker, G. Refael, J. I. Cirac, E. Demler, M. D. Lukin,
and P. Zoller, \href{https://journals.aps.org/prl/abstract/10.1103/PhysRevLett.106.220402}{Phys. Rev. Lett. \textbf{106},
220402 (2011).}

%\bb{JakZoller} \textbf{Creation of effective magnetic fields in optical lattices: the Hofstadter butterfly for cold neutral atoms}, D. Jaksch and P. Zoller, \href{https://iopscience.iop.org/article/10.1088/1367-2630/5/1/356}{New J. Phys. \textbf{5}, 56 (2003).}

\bb{AidelsGold} \textbf{Measuring the Chern number of Hofstadter bands with ultracold bosonic atoms}, M. Aidelsburger, M. Lohse, C. Schweizer, M. Atala, J. T.
Barreiro, S. Nascimb\'ene, N. R. Cooper, I. Bloch, and N.
Goldman, \href{https://www.nature.com/articles/nphys3171}{Nat. Phys. \textbf{11}, 162 (2015).}

\bb{Jotzu}\textbf{Experimental realization of the topological Haldane model with ultracold fermions}, G. Jotzu, M. Messer, R. Desbuquois, M. Lebrat, T.
Uehlinger, D. Greif, and T. Esslinger, \href{https://www.nature.com/articles/nature13915}{ Nature (London)
\textbf{515}, 237 (2014).}

\bb{Kapit_Mueller}\textbf{Optical-lattice Hamiltonians for relativistic quantum electrodynamics}, E. Kapit and E. Mueller,\href{https://journals.aps.org/pra/abstract/10.1103/PhysRevA.83.033625}{Phys. Rev. A \textbf{83}, 033625 (2011).}

\bb{Zoller_Ban} \textbf{Atomic Quantum Simulation of 
U(N) and SU(N) Non-Abelian Lattice Gauge Theories}, D. Banerjee, M. B\"ogli, M. Dalmonte, E. Rico, P. Stebler, U. J. Wiese and P. Zoller, \href{https://journals.aps.org/prl/abstract/10.1103/PhysRevLett.110.125303}{Phys. Rev. Lett. \textbf{110}, 125303 (2013).}

\bb{Wiese} \textbf{Ultracold quantum gases and lattice systems: quantum simulation of lattice gauge theories}, U.-J. Wiese, \href{https://onlinelibrary.wiley.com/doi/full/10.1002/andp.201300104}{Annalen der Physik \textbf{525}, 10 (2013).}

\bb{Zoller_Dal} \textbf{Atomic Quantum Simulation of Dynamical Gauge Fields Coupled to Fermionic Matter: From String Breaking to Evolution after a Quench}, D. Banerjee, M. Dalmonte, M. M\"uller, E. Rico, P. Stebler, U.-J. Wiese, and P. Zoller, \href{https://journals.aps.org/prl/abstract/10.1103/PhysRevLett.109.175302}{Phys. Rev. Lett. \textbf{109}, 175302 (2012).}

\bb{Lewenstein_Tag} \textbf{Simulation of non-Abelian gauge theories with optical lattices}, L. Tagliacozzo, A. Celi, P. Orland, M. W. Mitchell and M. Lewenstein, \href{https://www.nature.com/articles/ncomms3615}{Nature Commun. \textbf{4}, 2615 (2013).}

\bb{Zohar_Reznik}\textbf{Cold-Atom Quantum Simulator for SU(2) Yang-Mills Lattice Gauge Theory}, E. Zohar, J. I. Cirac, and B. Reznik, \href{}{Phys. Rev. Lett.
\textbf{110}, 125304 (2013).}


\bb{Baumann} \textbf{Dicke quantum phase transition with a superfluid gas in an optical cavity}, K. Baumann, C. Gulerin, F. Brennecke and T. Esslinger, \href{https://www.nature.com/articles/nature09009.pdf}{Nature \textbf{464}, 1301 (2010).}

\bb{PShakya}\textbf{Dimensional cross-over in self-organised super-radiant phases of ultra-cold atoms inside a cavity},
P. Shakya, A. Ratnakar, and S. Ghosh, \href{https://iopscience.iop.org/article/10.1088/1361-6455/acb119}{J. Phys. B: At. Mol. Opt. Phys. \textbf{56}, 035301 (2023))).}

\bb{Leonard1} \textbf{Supersolid formation in a quantum gas breaking a continuous translational symmetry}, J. L\'eonard, A. Morales, P. Zupancic, T. Esslinger, and
T. Donner, \href{https://www.nature.com/articles/nature21067.pdf}{Nature (London) \textbf{543}, 87 (2017).}


\bb{Kroeze}\textbf{Dynamical Spin-Orbit Coupling of a Quantum Gas},
R. M. Kroeze, Y. Guo, and B. L. Lev,
\href{https://journals.aps.org/prl/abstract/10.1103/PhysRevLett.123.160404}{Phys. Rev. Lett. \textbf{123}, 160404 (2019).}

\bb{Dreon}\textbf{Self-oscillating pump in a topological dissipative atom–cavity system},
D. Dreon, A. Baumg\"artner, X. Li, S. Hertlein, T. Esslinger and T. Donner, 
\href{https://www.nature.com/articles/s41586-022-04970-0}{Nature \textbf{608}, 494 (2022).}

\bb{Ballantine} \textbf{Meissner-like Effect for a Synthetic Gauge Field in Multimode Cavity QED}, K. E. Ballantine, B. Lev and J. Keeling, \href{https://journals.aps.org/prl/abstract/10.1103/PhysRevLett.118.045302}{Phys. Rev. Lett. \textbf{118}, 045302 (2017).}

\bb{Padhi2014} \textbf{Spin-orbit-coupled Bose-Einstein condensates in a cavity: Route to magnetic phases through cavity transmission}, B. Padhi and S. Ghosh, \href{https://journals.aps.org/pra/abstract/10.1103/PhysRevA.90.023627}{Phys. Rev. A \textbf{90}, 023627 (2014).}

\bb{Mivehvar2019}\textbf{Cavity-Quantum-Electrodynamical Toolbox for Quantum Magnetism}, F. Mivehvar, H. Ritsch, and F. Piazza, \href{https://journals.aps.org/prl/abstract/10.1103/PhysRevLett.122.113603}{Phys. Rev. Lett. \textbf{122}, 113603 (2019).}

\bb{Keeling2014} \textbf{Fermionic Superradiance in a Transversely Pumped Optical Cavity}, J. Keeling, M. J. Bhaseen and B. D. Simons, \href{https://journals.aps.org/prl/abstract/10.1103/PhysRevLett.112.143002}{Phys. Rev. Lett. \textbf{112}, 143002 (2014).}

\bb{Piazza2014} \textbf{Umklapp Superradiance with a Collisionless Quantum Degenerate Fermi Gas},
F. Piazza and P. Strack, \href{https://journals.aps.org/prl/abstract/10.1103/PhysRevLett.112.143003}{Phys. Rev. Lett. \textbf{112}, 143003 (2014).}

\bb{Chen2014} \textbf{Superradiance of Degenerate Fermi Gases in a Cavity}, Y. Chen, Z. Yu, and H. Zhai, \href{https://journals.aps.org/prl/abstract/10.1103/PhysRevLett.112.143004}{Phys. Rev. Lett. \textbf{112}, 143004 (2014).}

\bb{Pan}\textbf{Topological Superradiant States in a Degenerate Fermi Gas}, J.-S. Pan, X.-J. Liu, W. Zhang, W. Yi, and G.-C. Guo,
\href{https://journals.aps.org/prl/abstract/10.1103/PhysRevLett.115.045303}{Phys. Rev. Lett. 115, 045303 (2015).}

\bb{Zheng}\textbf{Synthetic topological
Kondo insulator in a pumped optical cavity}, Z. Zheng, X.-B. Zou, and G.-C. Guo,  \href{https://iopscience.iop.org/article/10.1088/1367-2630/aaaa50}{New J. Phys.
20, 023039 (2018).}

\bb{Brenn_Kollath}\textbf{Ultracold Fermions in a Cavity-Induced Artificial Magnetic
Field},  C. Kollath, A. Sheikhan, S. Wolff, and F. Brennecke, \href{https://journals.aps.org/prl/abstract/10.1103/PhysRevLett.116.060401}{Phys. Rev. Lett. \textbf{116}, 060401 (2016).}

\bb{Sheikh_Kollath} \textbf{Cavity-induced
chiral states of fermionic quantum gases}, A. Sheikhan, F. Brennecke, and C. Kollath, \href{https://journals.aps.org/pra/abstract/10.1103/PhysRevA.93.043609}{Phys. Rev. A \textbf{93},
043609 (2016).}

\bb{Wolff_Sheikh}\textbf{Dissipative time evolution of a chiral state after a quantum quench}, S. Wolff, A. Sheikhan, and C. Kollath, \href{https://journals.aps.org/pra/abstract/10.1103/PhysRevA.94.043609}{Phys. Rev. A \textbf{94}, 043609 (2016).}

\bb{Sheikh_Brenn}\textbf{Cavity-induced
generation of nontrivial topological states in a two-dimensional Fermi gas}, A. Sheikhan, F. Brennecke, and C. Kollath, \href{https://journals.aps.org/pra/abstract/10.1103/PhysRevA.94.061603}{Phys. Rev. A \textbf{94}, 061603(R) (2016).}

\bb{Halati_Sheikh}\textbf{Cavity-induced
artificial gauge field in a Bose-Hubbard ladder}, C.-M. Halati, A. Sheikhan, and C. Kollath, \href{https://journals.aps.org/pra/abstract/10.1103/PhysRevA.96.063621}{Phys. Rev. A \textbf{96}, 063621 (2017).}

\bb{Halati_Kollath}\textbf{Cavity-induced
spin-orbit coupling in an interacting bosonic wire,} C.-M. Halati, A. Sheikhan, and C. Kollath, \href{https://journals.aps.org/pra/abstract/10.1103/PhysRevA.99.033604}{Phys.
Rev. A \textbf{99}, 033604 (2019).}

\bb{RitschRev}\textbf{Cold atoms in cavity-generated dynamical optical potentials,} H. Ritsch, P. Domokos, F. Brennecke and T. Esslinger, \href{https://journals.aps.org/rmp/abstract/10.1103/RevModPhys.85.553}{ Rev. of Mod. Phys., \textbf{85}, 553 (2013)}.

\bb{RitschRev2}\textbf{Cavity QED with quantum gases: new paradigms in many-body physics,} F. Mivehvar, F. Piazza, T. Donner and H. Ritsch, \href{https://www.tandfonline.com/doi/full/10.1080/00018732.2021.1969727}{Adv. in Phys. \textbf{70}, 1 (2021)}.

\bb{Nayak2008}\textbf{Non-Abelian anyons and topological quantum computation}, C. Nayak, S. H. Simon, A. Stern, M. Freedman, and S. D. Sarma, \href{https://journals.aps.org/rmp/abstract/10.1103/RevModPhys.80.1083}{Rev. Mod. Phys. \textbf{80}, 1083 (2008).}

\bb{Stern2013}\textbf{Topological Quantum Computation—From Basic Concepts to First Experiments}, A. Stern and N. H. Lindner, \href{https://www.science.org/doi/10.1126/science.1231473}{Science \textbf{339}, 1179 (2013).}

\bb{Mueller} \textbf{Effect of a Nonuniform Magnetic Field on a Two-Dimensional Electron Gas in the Ballistic Regime}, J. E. M\"{u}ller, \href{https://journals.aps.org/prl/pdf/10.1103/PhysRevLett.68.385}{Phys. Rev. Lett. \textbf{68}, 385 (1992).}

\bb{Pdye} \textbf{Electrons in a Periodic Magnetic Field Induced by a Regular Array of Micromagnets}, P. D. Ye, D. Weiss, R. R. Gerhardts, M. Seeger, K. von Klitzing, K. Eberl and H. Nickel, \href{https://journals.aps.org/prl/abstract/10.1103/PhysRevLett.74.3013}{Phys. Rev. Lett. \textbf{74}, 3013 (1995).}

\bb{Marcus} \textbf{Snake States along Graphene 
p-n Junctions}, J. R. Williams and C. M. Marcus, \href{https://journals.aps.org/prl/abstract/10.1103/PhysRevLett.107.046602}{Phys. Rev. Lett. \textbf{107}, 046602 (2011).}

\bb{Rickhaus} \textbf{Snake trajectories in ultraclean graphene p-n junctions}, P. Rickhaus, P. Makk, M.-H. Liu, E. Tovari, M. Weiss, R. Maurand, K. Richter and C. Sch\"onenberger, \href{https://www.nature.com/articles/ncomms7470}{Nat. Comm. \textbf{6}, 6470 (2015).}

\bb{Peeters} \textbf{Snake orbits and related magnetic edge states}, J. Reijneirs and F. M. Peeters \href{https://iopscience.iop.org/article/10.1088/0953-8984/12/47/305}{J. Phys.:Condens. Matter \textbf{12}, 9771 (2000).}
 
\bb{Peeters1}\textbf{Confined magnetic guiding orbit states}, 
J. Reijniers, A. Matulis, K. Chang, F. M. Peeters and P. Vasilopoulos,
 \href{https://iopscience.iop.org/article/10.1209/epl/i2002-00189-8/pdf}
{Europhys. Lett., \textbf{59}(5), 749(2002).}

\bb{Nogaret} \textbf{Electron dynamics in inhomogeneous magnetic fields}, A. Nogaret, \href{https://iopscience.iop.org/article/10.1088/0953-8984/22/25/253201}{J. Phys.: Cond. Matter, \textbf{22}, 253201 (2010).}

\bb{Cserti}\textbf{Theory of snake states in graphene}, 
L. Oroszl\'any, P. Rakyta, A. Korm\'anyos, C. J. Lambert, and J. Cserti, 
\href{https://journals.aps.org/prb/pdf/10.1103/PhysRevB.77.081403}
{Phys. Rev. B, \textbf{77}, 081403(R)(2008).}

\bb{Egger}\textbf{Conductance quantization and snake states in graphene magnetic waveguides}, 
T. K. Ghosh, A. De Martino, W. H\"ausler, L. Dell'Anna, and R. Egger, 
\href{https://journals.aps.org/prb/pdf/10.1103/PhysRevB.77.081404}
{Phys. Rev. B \textbf{77}, 081404(R) (2008).}

\bb{Sim}\textbf{Magnetic edge states in graphene in non-uniform magnetic fields}, 
S. Park and H.-S. Sim, 
\href{https://journals.aps.org/prb/pdf/10.1103/PhysRevB.77.075433}
{Phys. Rev. B \textbf{77}, 075433 (2008).}

\bb{Nogaret1} \textbf{Crossover between magnetic and electric edges in quantum Hall systems}, A. Nogaret, P. Mondal, A. Kumar, S. Ghosh, H Beere and D. Ritchie, \href{https://journals.aps.org/prb/abstract/10.1103/PhysRevB.96.081302}{Phys. Rev. B \textbf{96}, 081302 (2017).} 

\bb{Puja} \textbf{Quantum transport through pairs of edge states of opposite chirality at electric and magnetic boundaries}, P. Mondal, A. Nogaret and S. Ghosh, \href{https://journals.aps.org/prb/references/10.1103/PhysRevB.98.125303}{Phys. Rev. B \textbf{98}, 125303 (2018)}.        
    
\bb{Watanabe} \textbf{Conductance oscillations induced by ballistic snake states in a graphene heterojunction}, T. Taychatanapat, J. Y. Tan, Y. Yeo, K. Watanabe, T. Taniguchi and B. \"Ozyilmaz, \href{https://www.nature.com/articles/ncomms7093}{Nat. Comms. \textbf{6}, Article number: 6093 (2015) }

\bb{Nogaret2} \textbf{Resistance Resonance Effects through Magnetic Edge States}, A. Nogaret, S. J. Bending, and M. Henini, \href{https://journals.aps.org/prl/abstract/10.1103/PhysRevLett.84.2231}{Phys. Rev. Lett. \textbf{84}, 2231(2000).}

\bb{Hara} \textbf{Transport in a two-dimensional electron-gas narrow channel with a magnetic-field gradient}, M. Hara, A. Endo, S. Katsumoto, and Y. Iye, \href{https://journals.aps.org/prb/abstract/10.1103/PhysRevB.69.153304}{Phys. Rev. B \textbf{69}, 153304 (2004).}

\bb{Solimany} \textbf{Electron in a magnetic quantum dot}, L. Solimany and B. Kramer, \href{https://www.sciencedirect.com/science/article/pii/0038109895004394}{Solid State Comm. \textbf{96} 471(1995).}

\bb{Sim1} \textbf{Magnetic edge states in a magnetic quantum dot}, H.-S. Sim, K.-H. Ahn, K. J. Chang, G. Ihm, N. Kim and S. J. Lee, \href{https://journals.aps.org/prl/abstract/10.1103/PhysRevLett.80.1501}{Phys. Rev. Lett. \textbf{80} 1501 (1998).}

\bb{Matulis} \textbf{Quantum states in a magnetic antidot}, J. Reijniers, F. M. Peeters and A. Matulis, \href{https://journals.aps.org/prb/abstract/10.1103/PhysRevB.59.2817}{Phys. Rev. B \textbf{59} 2817 (1999).}

\bb{Kim} \textbf{Electronic structure of a magnetic quantum ring}, N. Kim, G. Ihm, H.-S. Sim, and K. J. Chang, \href{https://journals.aps.org/prb/abstract/10.1103/PhysRevB.60.8767}{Phys. Rev. B \textbf{60} 8767 (1999).}    
    
%\bb{Mekhov2007} \textbf{Cavity-Enhanced Light Scattering in Optical Lattices to Probe Atomic Quantum Statistics}, I. B. Mekhov, C. Maschler, and H. Ritsch, \href{https://journals.aps.org/prl/abstract/10.1103/PhysRevLett.98.100402}{Phys. Rev. Lett. \textbf{98}, 100402 (2007).}
%
%\bb{Chen2007} \textbf{Cavity QED determination of atomic number statistics in optical lattices}, W. Chen, D. Meiser, and P. Meystre, \href{https://journals.aps.org/pra/abstract/10.1103/PhysRevA.75.023812}{Phys. Rev. A \textbf{75}, 023812 (2007).}
%
%\bb{Chen2009}\textbf{Cavity QED characterization of many-body atomic states in double-well potentials: Role of dissipation}, W. Chen and P. Meystre, \href{https://journals.aps.org/pra/abstract/10.1103/PhysRevA.79.043801}{Phys. Rev. A \textbf{79}, 043801 (2009).}
%
%\bb{Bhattacharjee2009} \textbf{Probing superfluidity of periodically trapped ultracold atoms in a cavity by transmission spectroscopy}, A. Bhattacherjee, T. Kumar, and M. Mohan, \href{https://doi.org/10.2478/s11534-009-0158-x}{Central European Journal of Physics \textbf{8}, 850 (2009).}
%
%\bb{Mekhov2012} \textbf{Quantum optics with ultracold quantum gases: towards the full quantum regime of the light–matter interaction}, I. B. Mekhov and H. Ritsch, \href{https://iopscience.iop.org/article/10.1088/0953-4075/45/10/102001/pdf}{J. Phys. B \textbf{45}, 102001 (2012).}
%
%\bb{Oztop2012} \textbf{Excitations of optically driven atomic condensate in a cavity: theory of photodetection measurements}, B. O\"ztop, M. Bordyuh, O. E. Mu\"stecaplolu, and H. E. Tr\"ueci, \href{https://iopscience.iop.org/article/10.1088/1367-2630/14/8/085011}{New J. Phys. \textbf{14}, 085011 (2012).}
%
%\bb{Kozlowski2015} \textbf{Probing matter-field and atom-number correlations in optical lattices by global nondestructive addressing}, W. Kozlowski, S. F. Caballero-Benitez, and I. B. Mekhov, \href{https://journals.aps.org/pra/abstract/10.1103/PhysRevA.92.013613}{Phys. Rev. A \textbf{92}, 013613 (2015).}
%
%\bb{Brennecke2013} \textbf{Real-time observation of fluctuations at the driven-dissipative Dicke phase transition}, F. Brennecke et al., \href{https://www.pnas.org/doi/full/10.1073/pnas.1306993110}{Proc. Natl. Acad. Sci. USA 110, 11763 (2013).}
%
%\bb{Landig2015} \textbf{Measuring the dynamic structure factor of a quantum gas undergoing a structural phase transition}, R. Landig, F. Brennecke, R. Mottl, T. Donner, and T. Esslinger, \href{https://www.nature.com/articles/ncomms8046}{Nature communications 6, 7046 (2015).}

\bb{Stamper-Kurn1}\textbf
{Observation of quantum-measurement backaction with an ultracold atomic gas}, 
K. W. Murch, K. L. Moore, S. Gupta and D. M. Stamper-Kurn ,
\href{https://www.nature.com/articles/nphys965.pdf}
{Nature Physics \textbf{4}, 561-564 (2008).}    
    
    
\bb{Folman} \textbf{Microscopic atom optics: from wires to an atom chip}, R. Folman, P. Kruger, J. Schmiedmayer, J. Denschlag, and C. Henkel, \href{https://arxiv.org/pdf/0805.2613.pdf}{Adv. At. Mol., Opt. Phys. 48, 263 (2002).}

\bb{Holland} \textbf{Atomtronics: Ultracold-atom analogues of electronic devices},  B. T. Seaman, M. Kr\"amer, D. Z. Anderson, and M. J. Holland , 
\href{ https://journals.aps.org/pra/abstract/10.1103/PhysRevA.75.023615 }{ Phys. Rev. A \textbf{75}, 023615 (2007).}

\bb{Amico} \textbf{Focus on atomtronics-enabled quantum technologies},  L. Amico, G. Birkl, M. Boshier, and L.-C. Kwek, 
\href{https://iopscience.iop.org/article/10.1088/1367-2630/aa5a6d/pdf}{ New J. Phys. \textbf{19}, 020201 (2017).}

\bb{DJaksch}\textbf{Optical lattices, ultracold atoms and quantum information processing}, D. Jaksch, \href{ https://doi.org/10.1080/00107510410001705486}{Contemporary Physics, \textbf{45}(5), 367(2004)}



\bb{Jaynes} \textbf{Comparison of Quantum and Semiclassical Radiation
Theories with Application to the Beam Maser }, E. T. Jaynes and F. W. Cummings, \href{https://ieeexplore.ieee.org/stamp/stamp.jsp?tp=&arnumber=1443594}{Proc. IEEE \textbf{51}, 89-109(1963).}

\bb{Born}\textbf{Zur Quantentheorie der Molekeln}, M. Born, R. Oppenheimer, \href{https://onlinelibrary.wiley.com/doi/10.1002/andp.19273892002}{Ann. Physik \textbf{84}, 457 (1930).}

\bb{Truhlar1} \textbf{On the determination of Born-Oppenheimer
nuclear motion wave functions including
complications due to conical intersections
and identical nuclei}, C.A. Mead, D.G. Truhlar, \href{https://aip.scitation.org/doi/pdf/10.1063/1.437734?class=pdf} {J. Chem. Phys. \textbf{70}, 2284 (1979).}

\bb{Berry} \textbf{Quantal phase factors accompanying adiabatic changes}, M.V. Berry, \href{https://royalsocietypublishing.org/doi/pdf/10.1098/rspa.1984.0023}{Proc. R. Soc. Lond. A \textbf{392}, 45 (1984).}

\bb{Wilczek} \textbf{Appearance of Gauge Structure in Simple Dynamical Systems}, F. Wilczek and A. Zee, \href{https://journals.aps.org/prl/pdf/10.1103/PhysRevLett.52.2111}{Phys. Rev. Lett. \textbf{25}, 2111 (1984).}

\bb{Moody} \textbf{\textcolor{black}{Realizations of Magnetic-Monopole Gauge Fields: Diatoms and Spin Precession}}, J. Moody, A. Shapere, and F. Wilczek, \href{https://journals.aps.org/prl/pdf/10.1103/PhysRevLett.56.893}{Phys. Rev. Lett. \textbf{56}, 893 (1986)}.

\bb{Mead} \textbf{Molecular Kramers Degeneracy and Non-Abelian Adiabatic Phase Factors}, C.A. Mead, \href{https://journals.aps.org/prl/pdf/10.1103/PhysRevLett.59.161}{Phys. Rev. Lett. \textbf{59}, 161 (1987).}

\bb{Shapere} \textit{Geometric Phases in Physics}, ed. A. Shapere, F. Wilczek World Scientific, Singapore (1989).

\bb{Sun} \textbf{High-order quantum adiabatic approximation and Berry's phase factor}, C.P. Sun, \href{https://iopscience.iop.org/article/10.1088/0305-4470/21/7/023}{J. Phys. A: Math. Gen. \textbf{21}, 1595 (1988).}

\bb{Sun1} \textbf{Generalizing Born-Oppenheimer approximations and observable effects of an induced gauge field}, C.P. Sun and M.L. Ge, \href{https://journals.aps.org/prd/pdf/10.1103/PhysRevD.41.1349}{Phys. Rev. D \textbf{41}, 1349 (1990).}

\bb{Haroche}\textbf{Trapping Atoms by the Vacuum Field in a Cavity}, S. Haroche, M. Brune and J. M. Raimond, \href{https://iopscience.iop.org/article/10.1209/0295-5075/14/1/004/pdf}{Europhys. Lett., \textbf{14}(l), 19 (1991).}

\bb{Lembessis}\textbf{Artificial gauge potentials for neutral atoms: an application in evanescent light fields}, V. E. Lembessis, \href{https://www.osapublishing.org/DirectPDFAccess/3749184E-DF53-9281-97B49CD8F3317F9B_286291/josab-31-6-1322.pdf?da=1&id=286291&seq=0&mobile=no}{J. Opt. Soc. Am. B \textbf{31}, 1322 (2014).}

\bb{Sacha} \textbf{Artificial magnetic field induced by an
evanescent wave}, M. Mochol and K. Sacha, \href{https://www.nature.com/articles/srep07672.pdf}{Sci. Rep.  \textbf{5}, 7672 (2015). }

\bb{Cheneau}  \textbf{Geometric potentials in quantum optics: A semiclassical interpretation}, M. Cheneau, S. P. Rath, T. Yefsah, K. J. G\"{u}nter, G. Juzeli\~unas and J. Dalibard, \href{https://iopscience.iop.org/article/10.1209/0295-5075/83/60001/pdf}{Eur. Phys. Lett. \textbf{83}, 60001(2008)}.

\bb{Landau} L. D. Landau and E. M. Lifshitz, {\it The classical theory of fields}, Chapter 3, Pergamon Press (1998).

\bb{Jackson} J. D. Jackson, {\it Classical Electrodynamics}, Wiley, New York (1962). 

\bb{Meystre1992}\textbf{Velocity dependent spontaneous emission: strong coupling regime}, P. Meystre, \href{https://www.sciencedirect.com/science/article/abs/pii/003040189290324K}{Optics Communication \textbf{90}, 41-45(1992).}

\bb{Kozlovskii2001}\textbf{Photon Emission of a two-level atom moving in a cavity}, A. Kozlovskii, \href{https://link.springer.com/article/10.1134/1.1410590}{Journal of Experimental and Theoretical Physics, \textbf{93}, 462-470 (2001).}

\bb{Qi2011}\textbf{Topological insulators and superconductors}, X.-L. Qi and S.-C. Zhang, \href{https://journals.aps.org/rmp/abstract/10.1103/RevModPhys.83.1057}{Rev. Mod. Phys. \textbf{83}, 1057 (2011).}

\bb{Seo2012} \textbf{Topological phase transitions in ultracold Fermi superfluids: The evolution from Bardeen-Cooper-Schrieffer to Bose-Einstein-condensate superfluids under artificial spin-orbit fields}, K. Seo, L. Han, and C. A. R. S\'a de Melo, \href{https://journals.aps.org/pra/abstract/10.1103/PhysRevA.85.033601}{Phys. Rev. A \textbf{85}, 033601 (2012).}

\bb{Sato2009}\textbf{Non-Abelian Topological Order in $s$-Wave Superfluids of Ultracold Fermionic Atoms}, M. Sato, Y. Takahashi, and S. Fujimoto, \href{https://journals.aps.org/prl/abstract/10.1103/PhysRevLett.103.020401}{Phys. Rev. Lett. \textbf{103}, 020401 (2009)}.

\bb{Wu2002}\textbf{Superfluidity of Bose–Einstein condensate in an optical lattice: Landau–Zener tunnelling and dynamical instability}, B. Wu, Q. Niu, \href{https://iopscience.iop.org/article/10.1088/1367-2630/5/1/104}{New J. Phys. \textbf{5}, 104 (2003).}

\bb{Ferris2008}\textbf{Dynamical instabilities of Bose–Einstein
condensates at the band edge in one-dimensional optical lattices}, A. J. Ferris, M. J. Davis, R. W. Geursen, P. B. Blakie, and A. C. Wilson, \href{https://journals.aps.org/pra/abstract/10.1103/PhysRevA.77.012712}{Phys. Rev. A \textbf{77}, 012712 (2008).}

\bb{Courteille}\textbf{Superradiant Rayleigh Scattering and Collective Atomic Recoil Lasing in a Ring Cavity}, S. Slama, S. Bux, G. Krenz, C. Zimmermann, 
and Ph.W. Courteille.  
\href{https://journals.aps.org/prl/pdf/10.1103/PhysRevLett.98.053603}
{Phys. Rev. Lett., \textbf{98}, 053603 (2007).}

\bb{Shore} \textbf{Is a quantum standing wave composed of two travelling waves}, B. W. Shore, P. Meystre and S. Stenholm, \href{https://www.osapublishing.org/DirectPDFAccess/2C64A143-C70E-6AE4-DB88A52A5130CF5F_5989/josab-8-4-903.pdf?da=1&id=5989&seq=0&mobile=no}{J. Opt. Soc. Am. B, \textbf{8}(4), 903 (1991)}.

\bb{Sakurai} J. J. Sakurai, \textit{Modern Quantum Mechanics, Revised Edition}, ed. S. F. Tuan, Addison-Wesley Publishing Company (1994).

\bb{Nicacio} \textbf{Coupled harmonic systems as quantum buses in
thermal environments}, F. Nicacio and F. L. Semiao, \href{https://iopscience.iop.org/article/10.1088/1751-8113/49/37/375303/pdf}{J. Phys. A: Math. Theor. \textbf{49}, 375303 (2016).} (Appendix A)


\end{thebibliography}
\end{document}